\documentclass[sigconf, natbib, screen]{acmart}

\acmSubmissionID{5390}

\usepackage{multirow}
\usepackage{mathtools}
\usepackage{svg}
\usepackage[font=small,skip=0pt]{caption}
\usepackage{graphicx}
\usepackage{subcaption}
\usepackage[inline]{enumitem}
\usepackage{booktabs}
\usepackage{adjustbox}
\usepackage{hyphenat}
\usepackage{acronym}
\usepackage{cleveref}[2012/02/15]
\crefformat{footnote}{#2\footnotemark[#1]#3}
\usepackage{pifont}

\usepackage{bigdelim}

\makeatletter
\renewcommand\paragraph{\@startsection{paragraph}{4}{\z@}%
                       {-2\p@ \@plus -1\p@ \@minus -1\p@}%
                       {-0.5em \@plus -0.22em \@minus -0.1em}%
                       {\normalfont\normalsize\itshape}}
\makeatother

\newcommand{\header}[1]{\vspace*{0.8mm}\noindent\textbf{#1.}}

\newcommand{\up}[1]{{#1}$^\blacktriangle$}
\newcommand{\down}[1]{{#1}$^\blacktriangledown$}
\newcommand{\pad}[1]{{#1}\phantom{$^\blacktriangledown$}}

\acrodef{L2R}{learning-to-rank}
\acrodef{IPS}{inverse propensity scoring}
\acrodef{SERP}{search result page}
\acrodef{nDCG}{normalized discounted cumulative gain}
\acrodef{PPL}{perplexity}
\acrodef{PDS}{policy distributional shift}
\acrodef{OPE}{off-policy evaluation}
\acrodef{OPS}{off-policy selection}
\acrodef{ind PPL}{in-distribution perplexity}
\acrodef{ood PPL}{out-of-distribution perplexity}
\acrodef{CMIP}{conditional mutual information with the logging policy}
\acrodef{CTR}{click-through rate}

\acrodef{MI}{mutual information}
\acrodef{CMI}{conditional mutual information}

\acrodef{DCTR}{document CTR model}
\acrodef{RDCTR}{ranked document CTR model}
\acrodef{UBM}{user browsing model}
\acrodef{PBM}{position-based model}
\acrodef{UBM}{user browsing model}
\acrodef{DBN}{dynamic Bayesian network model}
\acrodef{NCM}{neural click model}
\acrodef{CACM}{context-aware click model}

\allowdisplaybreaks

\begin{document}
\copyrightyear{2023}
\acmYear{2023}
\setcopyright{acmlicensed}
\acmConference[SIGIR '23]{Proceedings of the 46th International ACM SIGIR Conference on Research and Development in Information Retrieval}{July 23--27, 2023}{Taipei, Taiwan}
\acmBooktitle{Proceedings of the 46th International ACM SIGIR Conference on Research and Development in Information Retrieval (SIGIR '23), July 23--27, 2023, Taipei, Taiwan}
\acmPrice{15.00}
\acmDOI{10.1145/3539618.3591639}
\acmISBN{978-1-4503-9408-6/23/07}

\settopmatter{printfolios=false}

\ccsdesc[500]{Information systems~Query log analysis}

\keywords{Click models, Offline evaluation, Counterfactual learning-to-rank}

\author{Romain Deffayet}
\authornote{Both authors contributed equally to the paper}
\orcid{0000-0001-8265-9092}
\affiliation{%
\institution{Naver Labs Europe}
\city{Meylan}
\country{France}}
\affiliation{%
\institution{University of Amsterdam}
\city{Amsterdam}
\country{The Netherlands}}
\email{romain.deffayet@naverlabs.com}

\author{Philipp Hager}
\authornotemark[1]
\orcid{0000-0001-5696-9732}
\affiliation{%
\institution{University of Amsterdam}
\city{Amsterdam}
\country{The Netherlands}}
\affiliation{%
\institution{Booking.com}
\city{Amsterdam}
\country{The Netherlands}}
\email{p.k.hager@uva.nl}

\author{Jean-Michel Renders}
\orcid{0000-0002-7516-3707}
\affiliation{%
\institution{Naver Labs Europe}
\city{Meylan}
\country{France}}
\email{jean-michel.renders@naverlabs.com}

\author{Maarten de Rijke}
\orcid{0000-0002-1086-0202}
\affiliation{%
\institution{University of Amsterdam}
\city{Amsterdam}
\country{The Netherlands}}
\email{m.derijke@uva.nl}

\title{An Offline Metric for the Debiasedness of Click Models}

\begin{abstract}
A well-known problem when learning from user clicks are inherent biases prevalent in the data, such as position or trust bias. Click models are a common method for extracting information from user clicks, such as document relevance in web search, or to estimate click biases for downstream applications such as counterfactual \acl{L2R}, ad placement, or fair ranking. Recent work shows that the current evaluation practices in the community fail to guarantee that a well-performing click model generalizes well to downstream tasks in which the ranking distribution differs from the training distribution, i.e., under covariate shift. In this work, we propose an evaluation metric based on conditional independence testing to detect a lack of robustness to covariate shift in click models. We introduce the concept of debiasedness in click modeling and derive a metric for measuring it. In extensive semi-synthetic experiments, we show that our proposed metric helps to predict the downstream performance of click models under covariate shift and is useful in an off-policy model selection setting.\vspace{-1mm}
\end{abstract}

\maketitle
\acresetall

\section{Introduction}
\label{sec:intro}

Search and recommender systems aim to rank items in order of relevance to a given search query or user context~\cite{liu-2009-ltr}. Operational search engines have access to large logs of user behavior that are valuable sources for improving ranking systems~\cite{yandex,mslr,chapelle-2011-yahoo}. However, implicit user feedback in the form of clicks is well-known to be biased~\cite{pbm,joachims-2005-position-bias}. E.g., clicks can only occur on items exposed to users, introducing selection bias~\cite{ovaisi-2020-selection-bias,policy-aware}. Also, the rank at which a document is displayed greatly impacts the number of users seeing and clicking an item, leading to position bias~\citep{joachims-2005-position-bias,unbiased_joachims}. And trust bias arises when users rely on their search engine to place relevant documents at the top leading to clicks on top-ranked items regardless of their relevance~\cite{agarwal-2019-trust-bias,affine-estimator}.

\header{Click models}
Click models have a long history in web search for modeling user behavior by learning to predict how a user would interact with a given list of items~\citep{cm_book,pbm,dbn,ubm,ncm}. Click models explicitly model effects that impact a user's click decision, such as item relevance, position bias, or trust bias, and are thus a valuable tool for understanding users~\citep{pbm}, predicting ad clicks~\citep{zhu-2010-novel-click-model,chen-2012-position,mcmahan-2013-ad}, as offline evaluation metrics~\citep{cm-offline-metrics}, or estimating biases that the field of unbiased \acl{L2R} aims to mitigate~\cite{unbiased_joachims,affine-estimator,unbiased-online-offline}.

Commonly, two aspects of click models are evaluated~\cite{grotov-2015-comparative,cm_book}. First, a model's ability to accurately predict clicks is commonly measured using the perplexity of the model on a hold-out test set of clicks~\cite{ubm}. Second, if a model estimates document relevance, metrics such as nDCG or MRR can be computed using relevance annotations gathered by human experts~\cite{dbn}.
Recently, \citet{deffayet-2022} have shown that the current evaluation protocol of perplexity and nDCG does not guarantee that the best-performing model generalizes well to predicting clicks on unseen rankings. By simulating a variety of user behaviors on  rankings created by different ranking policies, the authors show that the best-performing models on one ranking policy are not guaranteed to perform well when presented with the same documents in a different order. This setting simulates a covariate shift in the ranking distribution (also called policy shift).

\header{Failure of generalization}
\citet{deffayet-2022} identify two cases where the current evaluation protocol breaks down. First, they find that biased click prediction methods can achieve high nDCG scores, especially when the policy that collected the click data is already near-optimal and tends to generate similar rankings. Picture the case in which all documents are ranked in order of relevance to the user. In this setting, position bias perfectly correlates with document relevance, and naive methods such as using a document's average \acf{CTR} as relevance and click prediction will lead to strong nDCG and perplexity scores. But this method fails to predict clicks on the inverted ranking in which the most relevant item is displayed at the bottom and is, thus, highly affected by position bias. In this case, predicting the average \acs{CTR} of a document as inferred from the original dataset is not a good prediction of user behavior on the inverted ranking, and the model fails to generalize. This model is not invariant under policy shift because it gives different predictions depending on the train rankings.

Second, the authors find that click model mismatch, a case in which the assumptions of the click models do not match the user behavior in the collected dataset, can lead to wrong conclusions about which models generalize well to unseen rankings. While \citet{deffayet-2022} evaluate a variety of click models and identify trends about which click models tend to generalize better, we still lack a principled approach to reliably select a click model from a set of candidates for deployment in downstream applications.

\header{Debiasedness in click modeling}
A common approach in the field is to consider that the role of click models and counterfactual learning-to-rank algorithms is to explicitly cancel the biases present in the training data, in order to avoid propagating them to downstream applications (e.g., ~\cite{sofa, debiasing-conversion}).
In this work, we formalize the notion of \emph{debiasedness} of a click model's predictions w.r.t.\ the logging policy. We formulate it as the concept that the inferred relevance of a newly trained click model should not correlate with the relevance predictions of the policy that was used to collect the training data, beyond the true relevance signal. First, we prove that a click model's relevance predictions must be debiased for its click predictions to be invariant under policy shift. We also connect this concept to consistency and unbiasedness in click modeling. Secondly, we present \acfi{CMIP}, a method based on conditional independence testing that measures the degree of debiasedness of relevance predictions with respect to the logging policy.

In our semi-synthetic experiments, we first reproduce the findings in \citep{deffayet-2022} on strong but narrow logging policies. Then, we verify, on a wide array of training configurations, that \acs{CMIP} helps to predict the performance of models outside of their training distribution. Lastly, we show that \acl{OPS} strategies based on \acs{CMIP} incur lower regret than those based on \acl{PPL} and \acs{nDCG} only.

\header{Contributions}
Our contributions can be summarized as follows:

\begin{itemize}[leftmargin=*,nosep]
    \item We introduce the notion of debiasedness of a click model and show that it is necessary for the invariance of click prediction under policy shift.
    \item We propose \acs{CMIP}, a metric using relevance annotations that measures the degree of debiasedness of a click model.
    \item We show in semi-synthetic experiments that \acs{CMIP} improves predicting the downstream performance of click models as well as the regret of off-policy model selection strategies.
\end{itemize}

\noindent To support the reproducibility of this work, we release the code for this paper\footnote{\href{https://github.com/philipphager/sigir-cmip}{https://github.com/philipphager/sigir-cmip}} and a standalone implementation of our metric.\footnote{\href{https://github.com/philipphager/cmip}{https://github.com/philipphager/cmip}} Below, we first introduce related work on click models and conditional independence testing (Section~\ref{sec:related-work}). Then, we present the current evaluation protocol for click models and its deficiencies (Section~\ref{sec:background}) before introducing the concept of debiasedness and our proposed metric (Section~\ref{sec:method}). We end by evaluating our metric in extensive semi-synthetic experiments (Section~\ref{sec:experimental-setup} and~\ref{sec:results}).

\section{Related Work}
\label{sec:related-work}

\vspace*{-1mm}
\subsection{Click models and their evaluation}

Click models emerged to model user behavior in web search~\citep{pbm, ubm, dbn, TACM}. Early methods use probabilistic graphical models to encode assumptions about user behavior in order to disentangle the influence of the presentation of a search result and its intrinsic relevance. The examination hypothesis, for example, introduced with the position-based model (PBM)~\citep{pbm}, assumes that the user examine and perceive the document as relevant in order to click on it. The cascade model~\citep{pbm} assumes that users browse results from top to bottom, click on the first relevant result and then leave the page. For an overview of common click models, see \citep{cm_book}.

More recently, click models based on neural architectures have emerged~\citep{ncm,aicm,cacm,csm,mcm,graphcm} to model more complex browsing behavior~\citep{csm} and user preferences across sessions~\citep{cacm,graphcm}. Neural click models typically also use more expressive representations of queries, documents, and other meta-data~\cite{cacm,csm,graphcm}. Combined with recent optimization techniques, these models enable efficient training on large-scale click logs. In this work, we use three click models originally proposed as probabilistic graphical models and implement them with current gradient-based optimization techniques. We also include two neural click models, and two baselines based on click statistics. These models are presented in Section~\ref{sec:models}.

Click models are commonly evaluated using the log-likelihood of clicks in a test set, measuring how well a model approximates the observed data~\cite{grotov-2015-comparative}. \citet{pbm} evaluate models by measuring cross-entropy. More widely used nowadays is perplexity, which measures how surprised a model is to observe a click on a given document and rank~\cite{ubm}. Another line of work compares predicted and actual click-through-rates (CTRs) on a test set, typically using MAE or RMSE~\cite{dbn,zhu-2010-novel-click-model,grotov-2015-comparative}. \citet{aicm} introduced distributional coverage, a metric quantifying whether the distribution of click sequences predicted by a model matches the true distribution of clicks. Moving beyond click prediction, \citet{dbn} evaluate the ranking performance of click models by computing retrieval metrics (e.g., MAP or nDCG) on an additional test set of human relevance annotations. We introduce perplexity, nDCG, and their limitations in Section~\ref{sec:click-modeling}.

\vspace*{-3mm}
\subsection{Conditional independence testing}

This work introduces the concept of debiasedness, which leverages a test for conditional independence~\cite{dawid-1979-ci}. Given a set of three random variables $X$, $Y$, and $Z$, conditional independence assesses if, given $Z$, knowing $X$ is helpful for predicting $Y$ (and vice versa). Conditional independence tests are widely applied in statistics and causal inference, e.g., to verify edges in Bayesian networks~\cite{koller-2009-pgm}, to discover causal graphs~\cite{pearl-2009-causality}, or for feature selection~\cite{koller-1996-toward}.

We use a non-parameteric conditional independence test for continuous random variables. Approaches include binning continuous variables to apply tests for discrete data~\cite{margaritis-2005-discrete}, reframing the problem as measuring the distance between two conditional densities~\cite{su-2007-ci}, estimating conditional mutual information~\cite{mukherjee-2020-ccmi}, or using kernel-based methods~\cite{fukumizu-2004-kernel-hilbert, doran-2014-permutation-kernel}. We use methods from \citep{sen-2017-ccit,mukherjee-2020-ccmi} to estimate conditional mutual information. Their approach is inspired by the use of model-powered independence testing~\cite{lopez-2016-classifier-two-sample,sen-2017-ccit,mukherjee-2020-ccmi}, reformulating statistical tests as supervised learning problems, which can be solved using standard classification or regression models. We  introduce the approach we follow in Section~\ref{sec:metric}.

\vspace*{-1mm}
\section{Background}
\label{sec:background}

\vspace*{-1mm}
\subsection{Notation and assumptions}

\textbf{Notation.}
Let $d \in D$ be a document. A ranking $y$ is an ordered list of documents: $y = \left[d_1, d_2, \dots, d_K\right]$. Note that $y$ is a list of length $K$ and $y[k]$ is the document displayed at position $k$. We retrieve the position of a document in $y$ using $\mathrm{rank}(d \mid y)$. A policy $\pi$ serves a ranking $y$ in response to a search query $q$. We consider stochastic ranking policies $\pi(y \mid q)$, which are probability distributions over rankings, given a query. 
For each ranking displayed to a user, we observe a vector of binary feedback $c$ of length $K$, with each entry denoting a click or no click on the displayed item: $c[k] \in \{0, 1\}$. Thus, our final dataset contains observations of a user query, the displayed ranking, and the recorded clicks: $\mathcal{D} = \{(q_i, y_i, c_i)\}^{N}_{i=1}$. The production ranker that collects this training dataset is commonly called the logging policy, which we denote as $\pi_l$.

\header{Assumptions}
We follow a common assumption in \ac{L2R}, that user clicks are a noisy and biased indicator of how relevant a document is to a given query~\cite{unbiased_joachims,harrie-unbiasedness}. We denote the relevance of an item to a query as $r(d, q)$. As explained before, clicks are usually influenced by bias factors such as the item's position or the user's trust in the system. Depending on the specific click model, bias factors can depend only on the position of a document or even on other documents in the same ranking. We refer to the vector of bias factors for documents in a given ranking as $o(y)$.

Our theory considers the family of click models that follow the structure of the examination hypothesis~\citep{pbm,cm_book}, which assumes that to be clicked, a document has to be observed by a user and deemed as relevant. In a general form, the examination hypothesis demands that relevance $r$ and bias factors $o$ factorize as:
\begin{equation}
\label{eq:EH}
    \begin{split}
        \forall k \in \{1, \dots, K\}, \;\; P(c[k] = 1 \mid y, q) = r(y[k], q) \times o(y)[k].
    \end{split}    
\end{equation}
This generic formulation of the examination hypothesis, also used in \citep{xpa}, can account for users observing an item based on its position or even based on the relevance of surrounding documents. For simplicity, we assume that $o(y)[k] > 0$, $\forall y \in \mathcal{D}, k \in \{1, \dots, K\}$, excluding any bias that leads to an item having no chance of being clicked~\cite{unbiased_joachims}, such as item selection bias. However, our work can be extended to this case. Lastly, our discussions below consider only a single query $q$ to simplify our notation. All statements can be extended to a setting with multiple queries.

\vspace*{-2mm}
\subsection{Evaluating click models}
\label{sec:click-modeling}

Click models are trained on an objective quantifying the quality of their click prediction. However, their primary goal, arguably, is to recover accurate estimates of the latent factors of user feedback. Hence, the literature on click models has adopted metrics for both of these objectives, respectively \acf{PPL} and \acf{nDCG}.

\paragraph{The click prediction quality on a test set}is measured by the \acl{PPL} at each rank $k$ and by the average \acl{PPL} over all ranks:
\begin{align}
\mathrm{PPL}@k &= 2^{-\frac{1}{N} \sum_{(y,c) \in \mathcal{D}} c[k] \log_2 \tilde{c}[k] + (1 - c[k]) \log_2 (1 - \tilde{c}[k])},\\
\mathrm{PPL} &= \frac{1}{K} \sum_{k=1}^K \mathrm{PPL}@k,
\end{align}
where $\tilde{c} = P(c \mid y)$ is a vector of click probabilities predicted by the model for a ranking $y$. Perplexity measures how surprised a model is to observe a given click behavior in the test set, given the model's parameters~\cite{ubm}. Perplexity is at least one and can be arbitrarily high. However, since a model predicting clicks at random has a perplexity of two, a realistic click model should achieve a perplexity between one and two~\cite{cm_book}.

\paragraph{The quality of relevance estimates $\tilde{r}$}for documents is measured by the ranking metric \acs{nDCG}, comparing predicted relevance scores against human annotations of relevance:
\begin{equation}
\begin{split}
\mathrm{nDCG} = \frac{\mathrm{DCG}(\tilde{y})}{\mathrm{DCG}(y^{\text{true}})},
\text{ with } \mathrm{DCG}(y) = \sum_{k = 1}^{K} \frac{2^{r(y[k])} - 1}{\log_2(k + 1)},
\end{split}
\end{equation}
where $y^{\text{true}} = \operatorname{arg\,sort}^\downarrow_{d \in y} r(d)$ and $\tilde{y} = \operatorname{arg\,sort}^\downarrow_{d \in y} \tilde{r}(d)$ are obtained by ranking documents in order of relevance, as predicted by human annotators and the click model, respectively.

These two metrics are complementary in the sense that \acl{PPL} quantifies the goodness-of-fit of the model to the logged data while \acs{nDCG} quantifies the quality of the rankings produced by the recovered relevance estimates. However, as we recall below, previous work has warned about the poor generalizability of these two metrics in many practical scenarios.
\begin{figure*}[t]
    \begin{center}
      \includegraphics[width=0.80\textwidth]{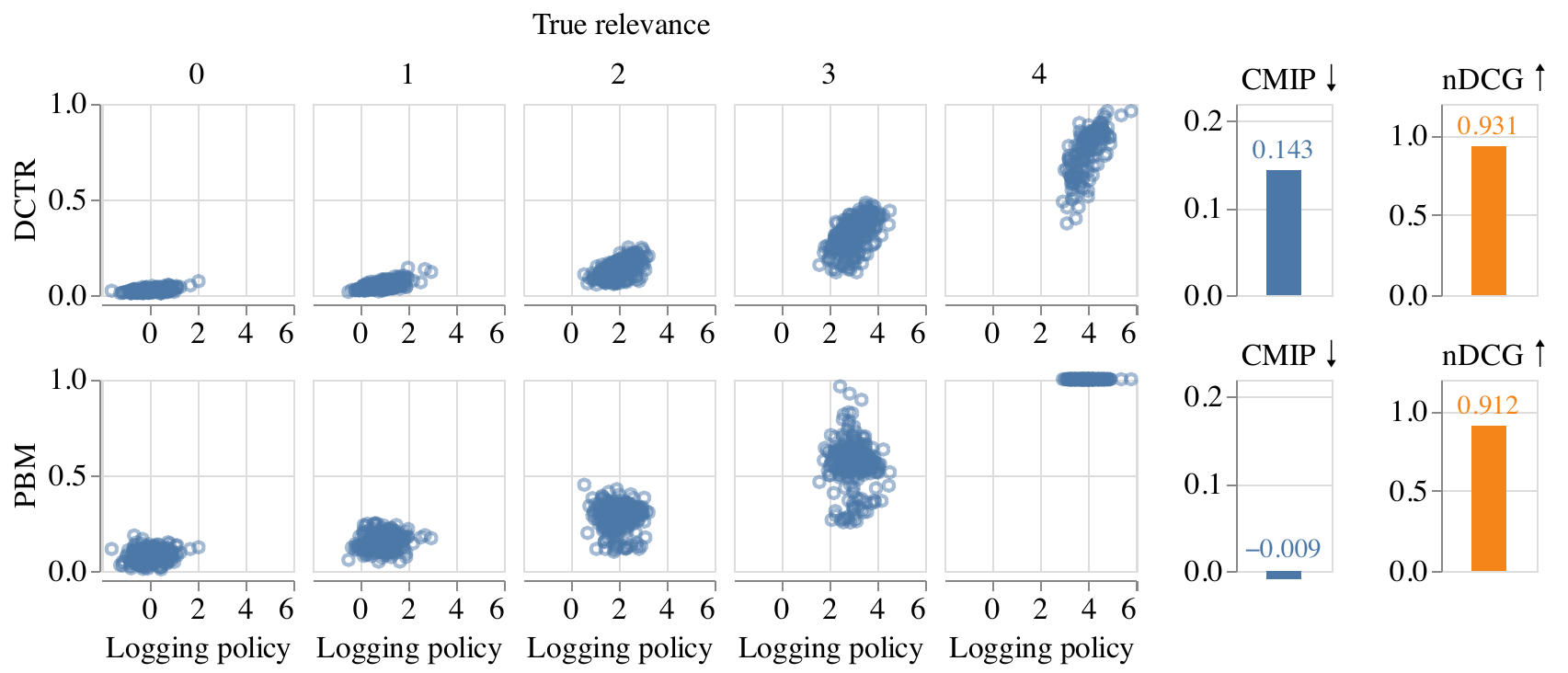}\\
      \caption{
        Comparing the relevance estimates of two click models (DCTR and PBM) against the relevance estimates of an almost optimal logging policy (NoisyOracle, defined in Section \ref{stochastic-policies}) for 1.5k documents, grouped by their true relevance. Clicks follow a PBM user model. The DCTR model achieves a higher nDCG but correlates notably with the logging policy, resulting in a high \acs{CMIP}. In contrast to the PBM, the DCTR model is not debiased in this setup. Note that \acs{CMIP} is in theory a non-negative metric but approximations can make it slightly negative.
      }
      \label{fig:cmip_dctr_pbm}
    \end{center}
\end{figure*}

\vspace*{-2mm}
\subsection{Perplexity fails to generalize, especially under model misfit}
\label{sec:ppl-fails}

Perplexity measures how well a model fits the conditional distribution of clicks given rankings observed in the dataset. However, the performance measured by perplexity only holds on a separate test set as long as the i.i.d.\ assumption is satisfied \cite{ood-generalization}, which notably requires that the rankings in the test set are sampled from the same distribution as rankings in the training set. This assumption is often violated when using click models to predict and evaluate ranking policies that differ from the one used for training (e.g.,~\citep{chen-2012-position, unbiased_joachims}). This mismatch creates a covariate shift in the input distribution of the model. In such cases, no guarantee can be derived on the out-of-distribution performance of the trained models, and we are likely to observe a drop in performance.

\citet{deffayet-2022} show the empirical effect of covariate shift on click model prediction. Model misfit, searching within model classes that do not contain the Bayes-optimal function, likely aggravates the performance drop on click predictions out-of-distri\-bution. Model misfit is not the only cause for the lack of robustness. E.g., deep neural networks, that have high capacity and potentially include the Bayes-optimal predictor, can also suffer from covariate shift due to training sets covering only a narrow space of all possible rankings. \emph{Hence, PPL measured in-distribution is not a good predictor of click model performance out-of-distribution in many practical scenarios, including model misfit and narrow logging policies.}

\vspace*{-2mm}
\subsection{nDCG fails to generalize when the logging policy is good}
\label{sec:ndcg-fails}

\acs{nDCG} assesses the ordering of documents based on their predicted relevance scores. As a list-wise metric, \acs{nDCG} does not evaluate the accuracy of the estimated relevance probabilities but only how ordering by these relevance estimates correlates with rankings obtained through expert annotations. This ranking task is a  use case for click models, but many scenarios require accurate estimation of relevance and examination probabilities, such as \acl{OPE}~\citep{oosterhuis-2020-taking}, counterfactual learning-to-rank with \acl{IPS}~\citep{agarwal-2019-trust-bias}, or click maximization in reinforcement learning~\citep{slivkins-2019-introduction}.
Performance measured by \acs{nDCG} can be misleading for these tasks since highly biased and poorly predictive click models can obtain high nDCG scores~\cite{deffayet-2022}. Indeed, when the logging policy already achieves a high nDCG, one cannot use nDCG to differentiate between a model predicting accurate relevance probabilities from a model replicating the logging policy, e.g., by sorting documents by their number of impressions. In more realistic scenarios, this misleading behavior of nDCG might manifest itself only for a group of queries (e.g., tail queries), enabling a model to achieve an improved nDCG score at the cost of biased relevance estimates for these queries. \emph{Consequently, \acs{nDCG} is not a good predictor of click model debiasing capabilities in many realistic settings.}

Faced with the lack of metrics evaluating the robustness of click models to shifts in the input rankings, we propose the idea of measuring debiasedness in the next section.

\section{Towards healthy benchmarks:\\ A metric to quantify debiasedness}
\label{sec:method}

To establish \acs{CMIP}, which measures the robustness of click models to covariate shift, we first formalize the notion of \emph{debiasedness} in Section \ref{sec:debiasedness}. We then explain how to test for this property in Section \ref{sec:cmi}, and finally instantiate our proposed metric in Section \ref{sec:metric}. 

\vspace*{-2mm}
\subsection{Debiasedness in click modeling}
\label{sec:debiasedness}

We first recall the data generation process in learning-to-rank, focusing on a single query to simplify our notation: we are given an i.i.d. sample $D$ of documents $d_i \sim \mathcal{P}(d)$ and a dataset $\mathcal{D}$ of i.i.d. rankings $y_j \sim \pi_l(y|D)$ and clicks $c_j \sim \mathcal{P}(c | y_j)$. Each document $d$ is assigned a true relevance score $r(d)$ and a relevance score estimated by the logging policy $r_l(d)$.\footnote{In practice, the true relevance score is often estimated by expert annotators and the the logging policy score is either known to the practitioner or can be estimated on a dataset $\mathcal{D}$ of rankings as $r_l(d) \approx \mathbb{E}_{y \sim \mathcal{D}} \left[ 1 / \mathrm{rank}(d | y) \right]$.}

\subsubsection{Debiasedness of a relevance scoring function}
In this work, we study how a candidate scoring function $\tilde{r} : d \mapsto \mathbb{R}$ relates to the true relevance scores. In particular, we look at the distribution of scores across documents by introducing the notion of debiasedness of a scoring function $\tilde{r}$. To do so, we first define the random variables $R$, $R_l$ and $\tilde{R}$ which map a document $d \sim \mathcal{P}(d)$ to their corresponding relevance scores $r(d)$, $r_l(d)$ and $\tilde{r}(d)$.

\begin{definition}
\label{def-debiasedness-scores}
A scoring function $\tilde{r}$ is \emph{debiased w.r.t.\ the logging policy} if its corresponding random variable $\tilde{R}$ is independent of the relevance of the logging policy $R_l$, conditioned on the true relevance~$R$:
\begin{equation}
\tilde{R} \Perp R_l \mid  R.
\end{equation}
\end{definition}

\noindent Intuitively, debiasedeness means that the score of a document cannot be predicted by knowing where the logging policy placed it. Click models aim to disentangle the factors influencing user behavior, thereby also alleviating biases induced by the logging policy. Therefore, a natural property that we may expect of a well-behaved click model is that its estimated relevance of an item cannot be predicted by revealing the relevance of that same item according to the logging policy, i.e., \emph{debiasedness}.

We give a visual intuition of debiasedness and our proposed metric \acs{CMIP} in Figure~\ref{fig:cmip_dctr_pbm}. We display the relevance estimates obtained by a PBM click model and a model predicting the average CTR of each document (DCTR) as relevance. In the plot, we group all documents by their true annotated relevance. In contrast to the \acs{PBM}, the \acs{DCTR} model does not account for position bias simulated in the click data. We can observe a clear correlation of the relevance estimates of the \acs{DCTR} model with those of the logging policy, meaning we can predict the estimated relevance of a randomly drawn document by knowing where the logging policy placed it. It is a strong indication that the scores recovered by the \acs{DCTR} model are not debiased in this setting, which is captured in a higher score of~\acs{CMIP}. 

Finally, note that debiasedness alone does not guarantee high performance of a scoring function, e.g., random relevance scores are trivially debiased.

\subsubsection{A debiasing click model}

We now analyze the properties of a model that yields debiased scoring functions, i.e., a debiasing model.

\begin{definition}
\label{def-debiasedness-CM}
A click model is \emph{debiasing w.r.t.\ the logging policy} if its estimated relevance after training is independent of the relevance of the logging policy, conditionally on the true relevance as well as the dataset it has been trained on:
\begin{equation}
\tilde{R}^\mathcal{D} \Perp R_l \mid  (R, \mathcal{D}).
\end{equation} 
\end{definition}

Definition \ref{def-debiasedness-CM} is very strict as it requires that the scores recovered on any dataset are debiased, and it will likely not be satisfied by usual click models in practice. However, we believe it can act as a north star as it is a necessary condition for invariance of the click prediction under policy shift:

\begin{definition}
\label{def-invariance}
\rm
A click model is said to be \emph{invariant under policy shift} if its estimated click probabilities are the same regardless of the training dataset, i.e., for every ranking $y$ and any two datasets $\mathcal{D}_1$ and $\mathcal{D}_2$:
\begin{equation}
        \tilde{c}^{\mathcal{D}_1}(y) = \tilde{c}^{\mathcal{D}_2}(y) =\tilde{c}(y),
\end{equation}
where $\tilde{c}^\mathcal{D}(\cdot)$ are the click predictions obtained after training a click model on $\mathcal{D}$.
\end{definition}

\noindent%
This definition allows us to introduce our main theorem that for every click model following the examination-hypothesis (Eq.~\ref{eq:EH}), \emph{the invariance of click prediction under policy shift requires the model to be debiasing}:

\begin{theorem}
\label{theorem-invariant-debiased}
A click model that is invariant under policy shift is debiasing. For every dataset $\mathcal{D}$ and ranking $y$:
\begin{equation}
\label{invariant-implies-debiased}
    \tilde{c}^\mathcal{D}(y) =  \tilde{c}(y)
    \Rightarrow \tilde{R}^\mathcal{D} \Perp R_l \; | \;  (R, \mathcal{D}).
\end{equation}
\end{theorem}

We defer the proof of this theorem to Appendix \ref{app:proof}. Now that we have connected the notion of debiasedness to invariance of the click prediction, we propose to connect it to common goals in click modeling: unbiasedness and consistency.

\subsubsection{Connection to common objectives in counterfactual learning-to-rank}
Unbiasedness has been introduced in different subfields as a common goal of counterfactual \acl{L2R} \citep{harrie-unbiasedness, unbiased_joachims}. Extending the meaning of unbiasedness from estimators to click models, \emph{an unbiased click model recovers the true relevance parameters for each document, in expectation over possible training datasets}. Using $\tilde{r}^\mathcal{D}(d)$ to denote the relevance of document $d$, as predicted by a click model after being trained on dataset $\mathcal{D}$, a click model is unbiased if, and only if:
\begin{equation}
    \mathbb{E}_{\mathcal{D} \sim \pi_l}\left[ \tilde{r}^{\mathcal{D}}(d) \right] = r(d)
\end{equation}
We can link unbiasedness with a weaker version of Definition~\ref{def-debiasedness-CM}:


\begin{corollary} In expectation over possible datasets, the scoring function recovered by an unbiased click model after training is debiased, i.e., 
\begin{equation}
\mathbb{E}_{\mathcal{D} \sim \pi_l}\left[\tilde{R}^\mathcal{D} \right] \Perp R_l \mid  R.
\end{equation} 
\end{corollary}

\begin{proof}
    Let $d \sim \mathcal{P}(d)$, $\mathbb{E}_{\mathcal{D} \sim \pi_l}\left[\tilde{R}^\mathcal{D}(d) \right] = r(d)$.
\end{proof}

Consistency has been introduced more recently \citep{harrie-unbiasedness}, as a more attainable goal for click models. \emph{A consistent click model recovers the true relevance parameters in the limit of infinite data:}
\begin{equation}
    \lim_{|\mathcal{D}| \rightarrow \infty} \tilde{r}^{\mathcal{D}}(d) = r(d)
\end{equation}
Similarly to unbiasedness, we can link consistency with a weaker version of Definition~\ref{def-debiasedness-CM}:

\begin{corollary} In the limit of infinite data, The scoring function recovered by a consistent click model after training is debiased, i.e., 
\begin{equation}
\lim_{\mathcal{D} \to \infty} \tilde{R}^\mathcal{D} \Perp R_l \mid  R.
\end{equation} \end{corollary}

\begin{proof}
    Let $d \sim \mathcal{P}(d)$, $\lim_{\mathcal{D} \to \infty} \tilde{R}^\mathcal{D}(d) = r(d)$.
\end{proof}

\medskip

In conclusion, while debiasedness alone does not guarantee that a click model inferred the correct parameters during training, it can be connected to consistency, unbiasedness, and invariance under policy shift. It constitutes an intermediary goal for designing robust click models and counterfactual learning-to-rank algorithms. 

This section was concerned with establishing theoretical definitions and properties for the idea of debiasing the training data. Having formulated it as enforcing a conditional independence relationship, we are now able to leverage the literature on conditional independence testing and propose an empirical procedure that tests for this property on a finite sample of data. We describe this procedure in the next section.

\vspace*{-2mm}
\subsection{Testing for debiasedness with mutual information}
\label{sec:cmi}

Given a set of annotated documents and the expected reciprocal rank of these documents under the logging policy, we might want to test for the debiasedness of a candidate click model's output using conditional independence testing. An independence test, given some significance level, would yield a binary answer to whether the predicted scores are debiased. In addition, it is unlikely that non-trivial models trained on finite data can ever be fully independent of their logging policy. In practice, however, we often have to pick a model from a set of candidates for deployment, i.e., \acf{OPS}. Therefore, we instead aim to quantify the degree of debiasedness using the effect size of an independence test: \acf{CMI}.

We first recall the concept of \acf{MI}, which measures the average reduction in uncertainty of a random variable $X$ obtained when knowing the value of a second random variable $Y$. 
\Acl{MI}, usually expressed as $I(X; Y)$, can capture non-linear relationships between variables.
Conditioning the mutual information between two variables on a third variable is strongly connected to conditional independence testing:
\begin{equation}
    \begin{split}
        \tilde{R} \Perp R_{l} \mid R \iff I(\tilde{R}; R_l \mid R) &= 0
    \end{split}
\end{equation}
where $\tilde{R}$ are the relevance scores predicted by a click model, $R_l$ the implicit relevance scores of the logging policy, and $R$ the human annotations of relevance. Meaning, conditional on $R$, knowing $R_l$ does not reduce the uncertainty of predicting $\tilde{R}$ and vice-versa. Thus, conditional independence testing is a special use case of \ac{CMI} and we can interpret a lower value of \ac{CMI} as a higher degree of debiasedness, with a \ac{CMI} of zero indicating that the relevance scores of a newly trained click model are independent of the policy that collected the dataset, conditional on the true relevance. We refer to the \acs{CMI} when computed w.r.t.\ the logging policy as \acs{CMIP}.

\vspace*{-2mm}
\subsection{Estimating \acf{CMIP}}
\label{sec:metric}

In this section, we cover how to estimate the \ac{CMIP} metric to quantify debiasedness. First, we note that \acs{CMI} can be expressed as the Kullback-Leibler divergence between two distributions:
\begin{equation}
    \begin{split}
        I(X; Y \mid Z) & = \mathcal{D}_{\mathrm{KL}} \left( p \| q  \right) \\
        &\text{with } p = P(\tilde{R}, R_l, R) \\
        & \text{ and } q = P(R) \; P(\tilde{R} \mid R) \; P(R_l \mid R)
    \end{split}
\end{equation}
which is a pseudo-distance between the joint distribution $p$ of all three variables occurring together and the distribution $q$ in which the predicted relevance scores $\tilde{R}$ and the relevance of the logging policies $R_l$ are independent, conditional on $R$. 

If the divergence between both distributions is zero, the joint distribution (which we actually observe) is equivalent to the distribution on which conditional independence holds. Given this divergence-based formulation of \ac{CMI}, we employ a two-step approach suggested in \citep{mukherjee-2020-ccmi}. First, we obtain samples from the marginal distribution $q$ on which conditional independence holds. Second, we estimate the KL-divergence between the observed dataset and the generated samples, which is the estimate of our \ac{CMIP} metric.

\vspace*{-1mm}
\subsubsection{Sampling from the marginal distribution $q$}

How can we obtain samples from the conditional independence distribution $q$ given our observational dataset $p$? 
For a proof that this methodology actually approximates $q$, we refer to \citep[Theorem 1]{sen-2017-ccit}.
We use a knn-based approach suggested in \citep{sen-2017-ccit}; its simplicity and computational speed make it suitable for an evaluation metric. Given a dataset of observed relevance labels for each document, $\mathcal{R} = \{ \left( \tilde{r}(d), r_{l}(d), r(d) \right) \}_{d \in \mathcal{D}}$, we split the data into two equal parts $\mathcal{R}_i$ and $\mathcal{R}_j$. For each document in $\mathcal{R}_i$, we find the nearest neighbor document in $\mathcal{R}_j$ with the most similar true relevance. In the case of using relevance annotations, this method simplifies to sampling any document from $\mathcal{R}_j$ with the same relevance label. By exchanging the relevance estimates of the logging policy between the two documents, the resulting dataset $\mathcal{R}_q = \{ (\tilde{r}(d_i), r_{l}(d_j), r(d_i))\}$, is now a sample from $q$.

\vspace*{-1mm}
\subsubsection{Estimating KL-divergence}
Given samples from the original relevance dataset $\mathcal{R}_p \sim p$ and samples from the marginal distribution $\mathcal{R}_q \sim q$, we can compute the \ac{CMI} as the KL-divergence between both distributions. We follow \citet{mukherjee-2020-ccmi} and frame the task of divergence estimation between two continuous joint distributions as a binary classification problem. The main idea is to label samples from $p$ with $m = 1$ and samples from $q$ with $m = 0$. After shuffling the two datasets into one, we train a binary classifier to predict to which distribution a given triplet of relevance values $\left( \tilde{r}, r_{l}, r \right)$ belongs to. The better the classifier can assign samples to their original distribution, the higher the divergence between the two distributions. Using the Donsker-Varadhan reformulation of KL-divergence~\citep[Definition 3]{mukherjee-2020-ccmi}, we use the classifier's predictions of $P(m = 1)$ on a test set to compute the \acfi{CMIP} as:
\begin{equation}
    \begin{split}
        \mbox{}\hspace*{-3mm}&\mathrm{CMIP} = \mathcal{D}_\mathrm{KL}\left(p \parallel q\right) \\
        \mbox{}\hspace*{-3mm}&\approx \frac{1}{|\mathcal{R}_p|} \!\sum_{i \in \mathcal{R}_p}\!{\log\frac{P(m = 1 \mid i)}{P(m = 0 \mid i)}} - \log \left(\frac{1}{|\mathcal{R}_q|} \!\sum_{j \in \mathcal{R}_q}\!{\frac{P(m = 1\mid j)}{P(m = 0\mid j)}} \right).
        \hspace*{-5mm}\mbox{}
    \end{split}
\end{equation}
The above procedure requires a well-calibrated classifier and we clip predictions $P(m = 1) \in  [0.01,  0.99]$ to avoid extremely large likelihood ratios when dividing by predictions close to zero. Lastly, we bootstrap the metric, performing multiple repetitions of k-nn sampling and KL divergence estimation, reporting the average over five repetitions. In order to simplify the usage of our metric, we release a standalone implementation of \acs{CMIP}.\footnote{\href{https://github.com/philipphager/cmip}{https://github.com/philipphager/cmip}}

\vspace*{-1mm}
\section{Experimental Setup}
\label{sec:experimental-setup}

We test if \ac{CMIP} helps to predict which click models are robust to covariate shift by performing experiments using a semi-synthetic click simulation setup prevalent in unbiased learning-to-rank~\cite{unbiased_joachims,policy-aware,affine-estimator,cm-ips}. The setup is semi-synthetic since we generate synthetic user clicks on real search queries and documents. To simulate shifts in the ranking distribution, we train models on click data collected under one logging policy and evaluate the model on clicks obtained under a different policy. Below, we introduce our simulation setup and the click models used in our experiments.

\vspace*{-2mm}
\subsection{Semi-synthetic click simulation}
\label{simulator}

\subsubsection{Overview}
We generate click datasets by repeatedly:
\begin{enumerate*}[label=(\roman*)]
    \item sampling a query and its candidate documents from a preprocessed real-world dataset;
    \item sampling a ranking of the candidate documents using a stochastic logging policy; and
    \item presenting the ranked search results to a synthetic user model to sample clicks.
\end{enumerate*}
In the following, we cover each step in more detail.

\vspace*{-1mm}
\subsubsection{Dataset and preprocessing}
\label{mslr-dataset-preprocessing}

Our click simulation is based on the MSLR-WEB10K dataset~\cite{mslr}. We use the training dataset of the first fold, containing 6,000 search queries, each with a set of candidate documents. Each query-document pair was judged by experts on a five point relevance scale: $r(d) \in \{0, 1, 2, 3, 4\}$, which we use as ground-truth in our experiments. During preprocessing, we reduce the number of documents per query to ten using stratified sampling on the human relevance annotation. Thereby, we reduce the number of candidate documents, while keeping a similar distribution of relevance grades. After discarding all queries with less than ten documents, we obtain a total of 5,888 queries.

\vspace*{-1mm}
\subsubsection{Stochastic policies}
\label{stochastic-policies}

For each new simulated user session, we first pick a query from the preprocessed dataset uniformly at random, a common practice in simulation for unbiased learning-to-rank~\cite{unbiased_joachims}, to avoid high variance on rare queries in this study. After sampling a query, we sample rankings of the candidate set of documents. For that, we first obtain relevance estimates for each document using one of three policies of different quality:

\begin{description}[nosep, style=unboxed, leftmargin=0cm]
    \item[Uniform:] A policy assigning the same relevance to all documents.
    \item[LambdaMART:] A LambdaMART ranker~\cite{lambdamart} trained on feature vectors and relevance annotations provided in MSLR-WEB10K.\footnote{LightGBM version 3.3.2, using 100 trees, 31 leafs, and learning rate 0.1.}
    \item[NoisyOracle:] A near-optimal policy using perturbed human relevance annotations after adding Gaussian noise of variance 0.5.
\end{description}

After using one of these three policies to obtain relevance estimates for each document, we sample stochastic rankings using a Plackett-Luce model~\cite{plackett,luce}. We sample multiple rankings per query to observe documents in different positions since a deterministic ranking would not allow our click models to disentangle effects such as position bias or relevance during training. We use the Gumbel Softmax trick to efficiently sample rankings from a Plackett-Luce distribution~\cite{bruch-2020-pl,harrie_plopt} and control the degree of stochasticity in the sampled rankings using the temperature parameter of the softmax.
We sample rankings with a low degree of stochasticity in our experiments, using a temperature of $T = 0.1$ by default.

\vspace*{-1mm}
\subsubsection{User models}
\label{user-model}
After sampling rankings, we generate synthetic clicks on our documents. We define how relevant each document is to the synthetic user based on the expert relevance annotations~\cite{err}: $R_d = \epsilon + \left(1 - \epsilon\right) \frac{2^{r(d)} - 1}{2^4 - 1}$, with noise $\epsilon = 0.1$ to also sample clicks on irrelevant documents. To examine our metric under a variety of click behaviors, we simulate four different users:

\begin{description}[nosep, style=unboxed, leftmargin=0cm]
    \item[PBM:] A user behaving according to the examination hypothesis, clicking only on observed and relevant documents. The observation probability depends only on the document position~\citep{pbm}. Following \citep{unbiased_joachims}, we define the observation probability at rank $k$ as: $O_k = \frac{1}{k}$.
    \item[DBN:] A user for whom relevance is split into two concepts: attractiveness and satisfaction. Attractiveness measures how likely a user is to click on a document after observing it in the ranking, while satisfaction estimates how likely a user is satisfied with the document after opening it. Documents are examined from top to bottom until the user is satisfied or abandons the list~\cite{dbn}. Thus, examination of a document not only depends on its rank, but also on the documents examined before. We use the probability of relevance $R_d$ to define the attractiveness of a document as $A_d = R_d$ and its satisfaction as $S_d = \frac{R_d}{2}$, so that even on near-optimal policies, fulfilling a user's information need sometimes requires more than one click.
    \item[MixtureDBN:] This setting simulates platforms presenting results horizontally, where users may not inspect the document in order~\citep{csm}. This mixture is composed of two DBN users: 70\% of the time, the user inspects the results in the usual order from first to last rank, but 30\% of the time, the user inspects the ranking in reverse order.
    \item[Carousel:] This last user simulates settings in which documents are grouped in vertical rows or carousels~\citep{rahdari2022carousel}. For this specific setting only, we use 25 documents per query instead of ten. The user chooses one of five carousels, according to a \acs{PBM}, where the relevance of a carousel is taken to be the average relevance of the five documents that compose it. Then the user clicks on documents within the chosen carousels according to a \acs{DBN}.
\end{description}

Next, we introduce the click models that we use, but we note that none fits the MixtureDBN and Carousel user behavior. We include these complex click behaviors to evaluate the usefulness of our metric under model mismatch.

\vspace*{-2mm}
\subsection{Click model overview}
\label{click-models}

We compare seven click models in our simulations; two are naive baselines using click statistics; five are prominent click models from the literature. In line with earlier studies, the models we compare do not input document features beyond the document's id. Below, we only summarize the main idea for each approach. For details, we refer to~\cite{deffayet-2022}, which we follow closely in our implementation.

\vspace*{-1mm}
\subsubsection{Naive baselines}
\begin{description}[nosep, style=unboxed, leftmargin=0cm]
    \item[\acs{DCTR}:] The \aclu{DCTR} uses the mean click-through-rate of a document as both click and relevance prediction. Since the CTR is averaged over all document positions, this model naively assumes that users examine all ranks equally.
    \item[\acs{RDCTR}:] The \aclu{RDCTR} predicts the mean click-through-rate of a document at a given rank as click probability. We follow \citep[Eq.~3]{deffayet-2022} and estimate relevance as the sum of a document's CTR at each rank, weighted by the inverse of the average CTR of all documents at the given rank.
\end{description}

In both methods, rarely examined documents can cause extreme click predictions, such as predicting a click probability of zero for a document that was never clicked. To mitigate predictions that lead to arbitrarily high perplexity values, we use the empirical Bayes method and initialize each prediction with Beta priors estimated on our training data as suggested in~\cite{dbn}.

\begin{table*}[t]
  \centering
  \caption{Adjusted $R^2$ score of predicting the out-of-distribution perplexity using combinations of in-distribution perplexity, nDCG, and CMIP using regression (higher is better). We mark significantly higher $\blacktriangle$ or lower performance $\blacktriangledown$ compared to using \emph{ind PPL, nDCG} at a significance level of $\alpha = 0.0001$. Adding \acs{CMIP} improves predictions of \acs{ood PPL}, achieving the best performance when combining all three metrics, and combinations including \acs{CMIP} are more consistent across different experimental setups and covariate shifts.}
  \label{tab:results}
  \setlength{\tabcolsep}{4pt}
\begin{tabular}{lllrrrrrrr}
  \toprule
  \bf User model  &  \bf Logging policy  &  \bf Test policy  &  ind PPL  &  nDCG  &  CMIP  &  \bf ind PPL, nDCG  &  CMIP, ind PPL  &  CMIP, nDCG  &  Joint \\
  \midrule
  \multirow{4}{*}{PBM} & \multirow{2}{*}{NoisyOracle} & LambdaMART & \pad{0.276} & \pad{0.256} & \pad{0.439} & \pad{0.395} & \up{0.938} & \up{0.972} & \up{0.963} \\
  & & Uniform & \pad{0.277} & \pad{0.346} & \up{0.725} & \pad{0.304} & \up{0.966} & \up{0.969} & \up{0.970} \\
  & \multirow{2}{*}{LambdaMART} & NoisyOracle & \down{0.369} & \pad{0.907} & \pad{0.904} & \pad{0.861} & \pad{0.883} & \up{0.962} & \up{0.944} \\
  & & Uniform & \down{0.347} & \up{0.965} & \pad{0.931} & \pad{0.889} & \up{0.974} & \up{0.977} & \up{0.975} \\
  \midrule
  \multirow{4}{*}{DBN} & \multirow{2}{*}{NoisyOracle} & LambdaMART & \down{0.235} & \up{0.978} & \down{0.517} & \pad{0.693} & \up{0.787} & \up{0.977} & \up{0.897} \\
  & & Uniform & \down{0.254} & \up{0.984} & \down{0.545} & \pad{0.688} & \up{0.821} & \up{0.983} & \up{0.963} \\
  & \multirow{2}{*}{LambdaMART} & NoisyOracle & \down{0.941} & \up{0.997} & \up{0.997} & \pad{0.974} & \pad{0.970} & \up{0.997} & \pad{0.983} \\
  & & Uniform & \down{0.958} & \up{0.999} & \up{0.999} & \pad{0.983} & \pad{0.982} & \up{0.999} & \pad{0.982} \\
  \midrule
  \multirow{4}{*}{MixtureDBN} & \multirow{2}{*}{NoisyOracle} & LambdaMART & \down{0.805} & \down{0.149} & \pad{0.843} & \pad{0.869} & \up{0.910} & \up{0.906} & \up{0.914} \\
  & & Uniform & \down{0.618} & \down{0.266} & \up{0.859} & \pad{0.767} & \up{0.884} & \up{0.909} & \up{0.880} \\
  & \multirow{2}{*}{LambdaMART} & NoisyOracle & \pad{0.985} & \down{0.167} & \down{0.625} & \pad{0.988} & \pad{0.990} & \down{0.862} & \pad{0.985} \\
  & & Uniform & \down{0.951} & \down{0.167} & \down{0.735} & \pad{0.987} & \pad{0.984} & \down{0.902} & \pad{0.987} \\
  \midrule
  \multirow{4}{*}{Carousel} & \multirow{2}{*}{NoisyOracle} & LambdaMART & \pad{0.974} & \down{0.811} & \up{0.993} & \pad{0.970} & \pad{0.977} & \pad{0.978} & \pad{0.973} \\
  & & Uniform & \pad{0.954} & \pad{0.941} & \up{0.994} & \pad{0.947} & \pad{0.968} & \up{0.995} & \pad{0.942} \\
  & \multirow{2}{*}{LambdaMART} & NoisyOracle & \pad{0.993} & \down{0.129} & \down{0.688} & \pad{0.990} & \pad{0.991} & \down{0.773} & \pad{0.994} \\
  & & Uniform & \pad{0.993} & \down{0.113} & \down{0.722} & \pad{0.996} & \pad{0.994} & \down{0.759} & \pad{0.998} \\
  \midrule
  \multicolumn{3}{c}{\rule[0.5ex]{1in}{0.4pt} \textbf{Average} \rule[0.5ex]{1in}{0.4pt}} & \pad{0.683} & \pad{0.574} & \pad{0.782} & \pad{0.831} & \pad{0.939} & \pad{0.933} & \pad{0.959} \\
  \bottomrule
  \end{tabular}
\end{table*}

\vspace*{-1mm}
\subsubsection{Click models}
\label{sec:models}
We implement a \textbf{\acs{PBM}} and a \textbf{\acs{DBN}} click model matching the user behaviors introduced in Section~\ref{user-model}. In addition, we implement three other models:

\begin{description}[nosep, style=unboxed, leftmargin=0cm]
    \item[\acs{UBM}:] Extending the \ac{PBM}, the \aclu{UBM} assumes that examining an item depends in addition to its position also on the position of the latest clicked document~\cite{ubm}.
    \item[\acs{NCM}:] The \aclu{NCM} uses an RNN to iterate over the list of documents and predicts clicks at every step. While the model only predicts clicks and does not explicitly model relevance, \citet{ncm} suggest to use the click probability of an item when placed on top of a ranking as its relevance.
    \item[\acs{CACM}:] We implement another RNN-based model iterating over rankings; instead of predicting clicks it predicts the user's probability of examination at each rank. The resulting examination probability is multiplied with an estimated relevance probability to obtain the click prediction. The model is a variant of the \aclu{CACM} introduced in \citep{cacm} as proposed in \citep[Eq.~14]{deffayet-2022}. 
\end{description}

All models are implemented using PyTorch, and are trained by minimizing a binary cross-entropy loss between the predicted clicks and the observed clicks in the training dataset. Further implementation details are openly accessible in our code.\footnote{\href{https://github.com/philipphager/sigir-cmip}{https://github.com/philipphager/sigir-cmip}}

\vspace*{-2mm}
\subsection{Experiments}
In our experiments, we evaluate whether \acs{CMIP} helps to predict the performance of click models under covariate shift. Therefore, we first generate 5M training, 1M validation, and 1M test clicks on a strong baseline policy (LambdaMART or NoisyOracle). We use the training/validation sets to train models and the test set to compute the \acf{ind PPL}. We simulate a covariate shift with a second test set of 1M clicks generated by a different policy, called test policy, and report the \acf{ood PPL}. Lastly, we use the human-annotated relevance labels from the MSLR-WEB10K dataset to compute nDCG and \acs{CMIP}.

\begin{figure}[t]
    \includegraphics[width=0.38\textwidth]{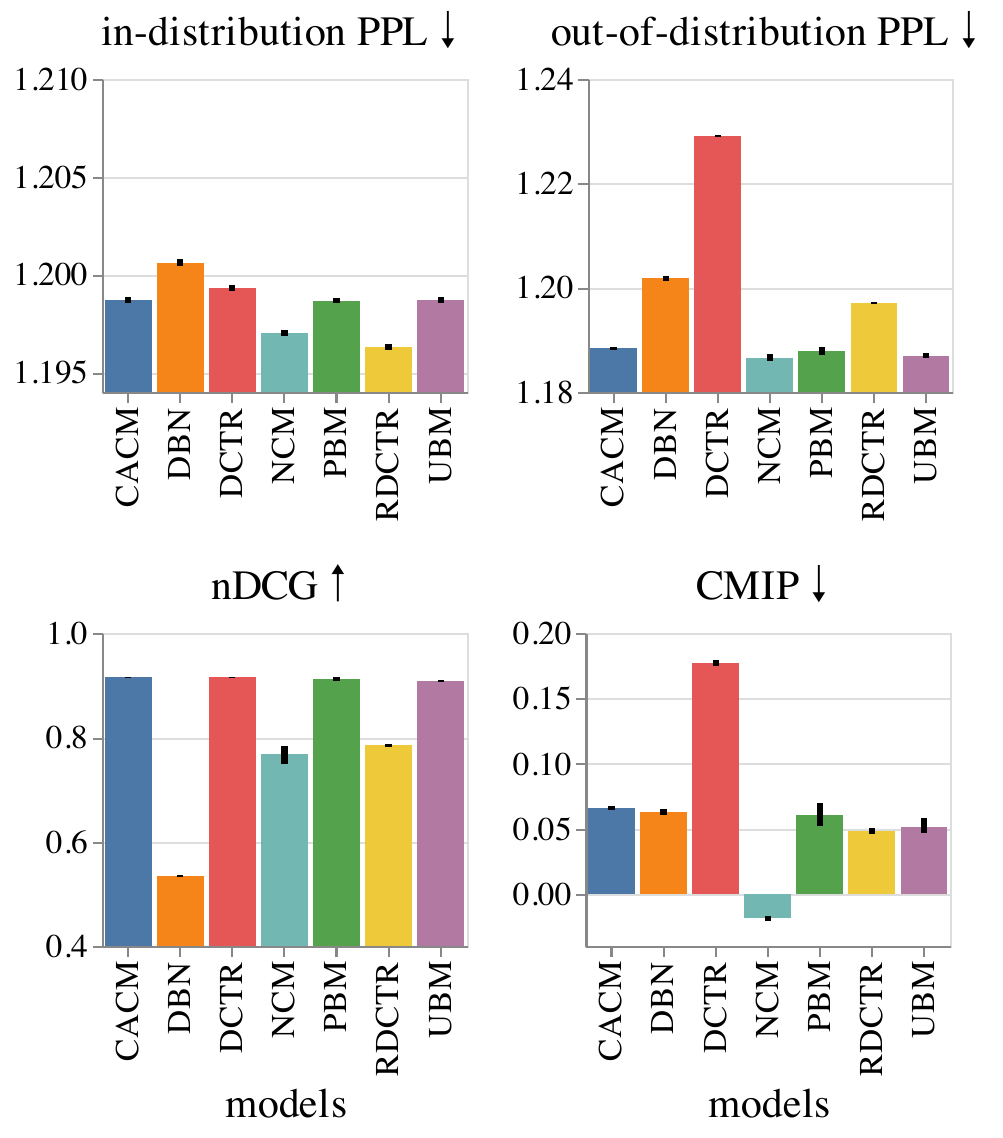}
    \caption{Comparing the performance of click models. Our proposed metric, \acs{CMIP}, helps predict out-of-distribution results. All models are trained on a PBM user model and a NoisyOracle logging policy and evaluated under a uniform policy. We average results over ten independent runs and we display the 95\% confidence interval.}
    \label{fig:visual-example}
\end{figure}

\vspace{-1mm}
\section{Results}
\label{sec:results}

\vspace*{-1mm}
\subsection{Evaluation with \acs{CMIP}: A visual example}
\label{sec:visual-results}

We introduce our experimental results by giving a visual intuition of \acs{CMIP}. In Figure~\ref{fig:visual-example}, we use a near-optimal logging policy (NoisyOracle) and generate clicks according to a PBM user model. We train seven click models and observe their \acf{ind PPL}, \acs{nDCG}, and our proposed metric \acs{CMIP}, as well as the models' performance under policy shift as measured by the \acf{ood PPL}.
We can see that neither \acs{ind PPL} nor \acs{nDCG} are sufficient to predict the downstream performance in \acs{ood PPL}. E.g., the competitive nDCG score of the DCTR model fails to capture the model's poor generalizability. On the contrary, \acs{CMIP} correctly identifies this model as biased by assigning it a high value. Similarly, despite a low \acs{nDCG}, NCM retains a good performance out-of-distribution as indicated by a low debiasedness score.

This visual example helps understand how \acs{CMIP} can be generally useful to evaluate click models, but the results are dependent on the configuration used for training and evaluating the models. Below, we  assess the predictive power of \acs{CMIP} more systematically.

\vspace*{-2mm}
\subsection{\acs{CMIP} helps predict out-of-distribution perplexity}
\label{predicts-ab}

Next, we systematically evaluate whether adding \acs{CMIP} to the existing metrics (\acs{ind PPL} and \acs{nDCG}) helps to predict the \acl{ood PPL} of click models. To do so, we test click models in a total of 16 configurations, where the user model, logging policy, and test policy vary. Moreover, we evaluate each model configuration over ten independent runs of our click simulation.

To quantify how well \acs{ind PPL}, \acs{nDCG}, \acs{CMIP}, and their combinations predict ood PPL, we use the metrics as input to a decision tree regressor predicting \acs{ood PPL}. Since the metric ranges depend on the policy and user model, we train separate regressors for each of the 16 user/policy configurations and seven metric combinations. We report predictive performance using the adjusted $R^2$ score on 2-fold cross validation with a thousand repetitions. We test differences between each metric combination and the current evaluation protocol of jointly using \acs{nDCG} and ind PPL using a two-tailed Welch's t-test with a significance level of $\alpha = 0.0001$, accounting for multiple comparisons using the Bonferroni correction.

We report the results in Table~\ref{tab:results}. First, neither \acs{ind PPL}, \acs{nDCG}, nor \acs{CMIP} are sufficient on their own to predict the \acs{ood PPL} well, even though \acs{CMIP} has a better mean predictive power than the other two metrics across all configurations. Second, using a combination of multiple metrics leads to better predictions of ood performance. However, we can observe that combining \acs{CMIP} with any other of the two metrics leads to an average $R^2$ greater than $0.9$ which is not attainable just using \acs{ind PPL} and \acs{nDCG}. More importantly, by inspecting the performance across configurations, we see that metric combinations including \acs{CMIP} are notably more consistent across different configurations, with their $R^2$ score rarely dropping below $0.8$. This suggests that \acs{CMIP} improves the safety and reliability of click model evaluation for deployments in downstream tasks. Lastly, the joint usage of all three metrics is either significantly better or on-par with the usage of \acs{nDCG} and \acs{ind PPL}. These trends are consistent when using linear regression and other regression metrics such as MSE. Our results strongly indicate that adding \acs{CMIP} to click modeling benchmarks should lead to more reliable predictions of downstream performance, and therefore help practitioners to mitigate the risks of deploying policies based on click models.

\begin{table}[t]
\caption{Average regret incurred by different \acs{OPS} strategies (lower is better). The regret is the difference in out-of-distribution click prediction performance with the best-performing model. We mark the best strategy in bold and underline the second best, and we report $95\%$ confidence intervals inside parentheses.}
\label{tab:regret}
\begin{tabular}{lc}
\toprule
Policies & Average regret \\
\midrule
PPL$\downarrow$ & $2.5499\; (\pm 0.0506)$ \\
nDCG$\uparrow$ & $5.2406\; (\pm 0.8480)$ \\
top-4 nDCG: PPL$\downarrow$& $1.6423\; (\pm 0.1417)$ \\
top-4 PPL: nDCG$\uparrow$ & $2.5199\; (\pm 0.3168)$  \\
\midrule
CMIP$\downarrow$ &  $2.526 8\; (\pm 0.2722)$ \\
top-4 CMIP: PPL$\downarrow$ & $2.5505\; (\pm 0.0507)$ \\
top-4 CMIP: nDCG$\uparrow$ & $\underline{0.9493}\; (\pm 0.1175)$ \\
top-4 CMIP, top-4 nDCG: PPL$\downarrow$ & $1.6404\; (\pm 0.1430)$ \\
top-4 CMIP, top-4 PPL: nDCG$\uparrow$ & $\mathbf{0.9176}\; (\pm 0.1221)$  \\
\bottomrule
\end{tabular}
\end{table}

\vspace*{-2mm}
\subsection{Strategies based on \acs{CMIP} incur lower regret in Off-Policy Selection Problems}

Next, we use \acs{CMIP} in an \acf{OPS} problem where we have a set of candidate click models and need to decide which one to use for downstream applications. To quantify how \acs{CMIP} helps practitioners select the best model, we design simple \acs{OPS} strategies based on the three metrics and compare the amount of regret they incur, i.e., how much click prediction performance, measured by \acs{ood PPL}, is lost by following a given strategy instead of selecting the optimal model. Every selection strategy is based on the maximization or minimization of a metric among the set of candidates. E.g., ``nDCG$\uparrow$'' is the strategy that selects the model with the highest nDCG. In addition to the three basic strategies defined this way, we define conditional strategies: e.g., ``top-4 \acs{CMIP}: PPL$\downarrow$'' selects the model with the lowest perplexity among the four models with the lowest \acs{CMIP}. The strategy ``top-4 \acs{CMIP}, top-4 PPL: nDCG$\uparrow$'' selects the model with the highest \acs{nDCG} among the intersection of the four models with lowest \acs{CMIP} and the four models with the lowest \acs{PPL}. If this intersection is empty, the model with highest nDCG is selected.

Table \ref{tab:regret} reports the average regret incurred by these strategies over the same configurations of user model, logging policy, and test policy as in Table \ref{tab:results}. For better readability, we multiply the obtained regret in terms of difference in \acs{ood PPL} by a factor $1000$. Also, the regret in \acs{OPS} is very sensitive to the exact set of candidates in the comparison, so in order to obtain more robust results, we apply each strategy on all possible combinations of five, six, or seven models from our set of seven candidates and report the average regret over these combinations.
We observe that most strategies based on \acs{CMIP} outperform those without it, and that the lowest average regret is obtained by ``top-4 \acs{CMIP}, top-4 PPL: nDCG$\uparrow$'', confirming that \acs{CMIP} is useful in \acl{OPS} problems.

\vspace*{-1mm}
\section{Conclusion}
\label{sec:conclusion}

We propose \acfi{CMIP}, an evaluation metric for click modeling benchmarks in unbiased \acl{L2R}. \ac{CMIP} addresses the problem that existing metrics do not ensure that click models are robust to shifts in the ranking policy and therefore fail to predict their performance on downstream tasks. \ac{CMIP} evaluates how relevance scores of trained models correlate with relevance scores of the logging policy beyond the true relevance signal, i.e., measuring how biased a new model is by the model that collected the training data.

\header{Findings and broader impact} We gave visual interpretations of \acs{CMIP} and its use for selecting best-performing click models. We quantified its usefulness in click modeling benchmarks by showing that it \begin{enumerate*}[label=(\roman*)] \item  improves the prediction of downstream performance when coupled with existing metrics, and \item lowers the regret incurred by \acl{OPS} strategies.\end{enumerate*} The effectiveness of \ac{CMIP} suggests that distributional approaches to offline evaluation, i.e., that consider the distribution of model outputs instead of individual predictions, may be useful to derive generalization properties. 

\header{Limitations and future work} First, \acs{CMIP} uses pointwise relevance annotations, but pairwise or listwise annotations could also be used. Second, we have assumed that annotations are a perfect predictor of relevance. It remains unclear how to interpret \acs{nDCG} and \acs{CMIP} in case annotator disagreement and biases render annotations less reliable. Finally, click feedback collected on fully randomized rankings could replace the need for expert annotations; we leave an analysis of \acs{CMIP} in that case for future work.


\vspace*{-2mm}
\begin{acks}
We thank Onno Zoeter and Harrie Oosterhuis for insightful discussions and comments.
This research was (partially) funded by the Hybrid Intelligence Center, a 10-year program funded by the Dutch Ministry of Education, Culture and Science through the Netherlands Organisation for Scientific Research, \url{https://hybrid-intelligence-centre.nl}.
All content represents the opinion of the authors, which is not necessarily shared or endorsed by their respective employers and/or sponsors.
\end{acks}

\bibliographystyle{ACM-Reference-Format}
\balance
\bibliography{main}


\begin{thebibliography}{54}


\ifx \showCODEN    \undefined \def \showCODEN     #1{\unskip}     \fi
\ifx \showDOI      \undefined \def \showDOI       #1{#1}\fi
\ifx \showISBNx    \undefined \def \showISBNx     #1{\unskip}     \fi
\ifx \showISBNxiii \undefined \def \showISBNxiii  #1{\unskip}     \fi
\ifx \showISSN     \undefined \def \showISSN      #1{\unskip}     \fi
\ifx \showLCCN     \undefined \def \showLCCN      #1{\unskip}     \fi
\ifx \shownote     \undefined \def \shownote      #1{#1}          \fi
\ifx \showarticletitle \undefined \def \showarticletitle #1{#1}   \fi
\ifx \showURL      \undefined \def \showURL       {\relax}        \fi
\providecommand\bibfield[2]{#2}
\providecommand\bibinfo[2]{#2}
\providecommand\natexlab[1]{#1}
\providecommand\showeprint[2][]{arXiv:#2}

\bibitem[Agarwal et~al\mbox{.}(2019)]%
        {agarwal-2019-trust-bias}
\bibfield{author}{\bibinfo{person}{Aman Agarwal}, \bibinfo{person}{Xuanhui
  Wang}, \bibinfo{person}{Cheng Li}, \bibinfo{person}{Michael Bendersky}, {and}
  \bibinfo{person}{Marc Najork}.} \bibinfo{year}{2019}\natexlab{}.
\newblock \showarticletitle{Addressing Trust Bias for Unbiased
  Learning-to-Rank}. In \bibinfo{booktitle}{\emph{The World Wide Web
  Conference}} (San Francisco, CA, USA) \emph{(\bibinfo{series}{WWW '19})}.
  \bibinfo{publisher}{Association for Computing Machinery},
  \bibinfo{address}{New York, NY, USA}, \bibinfo{pages}{4–14}.
\newblock
\showISBNx{9781450366748}
\urldef\tempurl%
\url{https://doi.org/10.1145/3308558.3313697}
\showDOI{\tempurl}


\bibitem[Ai et~al\mbox{.}(2021)]%
        {unbiased-online-offline}
\bibfield{author}{\bibinfo{person}{Qingyao Ai}, \bibinfo{person}{Tao Yang},
  \bibinfo{person}{Huazheng Wang}, {and} \bibinfo{person}{Jiaxin Mao}.}
  \bibinfo{year}{2021}\natexlab{}.
\newblock \showarticletitle{Unbiased Learning to Rank: Online or Offline?}
\newblock \bibinfo{journal}{\emph{ACM Trans. Inf. Syst.}} \bibinfo{volume}{39},
  \bibinfo{number}{2}, Article \bibinfo{articleno}{21} (\bibinfo{date}{feb}
  \bibinfo{year}{2021}), \bibinfo{numpages}{29}~pages.
\newblock
\showISSN{1046-8188}
\urldef\tempurl%
\url{https://doi.org/10.1145/3439861}
\showDOI{\tempurl}


\bibitem[Borisov et~al\mbox{.}(2016)]%
        {ncm}
\bibfield{author}{\bibinfo{person}{Alexey Borisov}, \bibinfo{person}{Ilya
  Markov}, \bibinfo{person}{Maarten de Rijke}, {and} \bibinfo{person}{Pavel
  Serdyukov}.} \bibinfo{year}{2016}\natexlab{}.
\newblock \showarticletitle{A Neural Click Model for Web Search}. In
  \bibinfo{booktitle}{\emph{Proceedings of the 25th International Conference on
  World Wide Web}} (Montr\'{e}al, Qu\'{e}bec, Canada)
  \emph{(\bibinfo{series}{WWW '16})}. \bibinfo{publisher}{International World
  Wide Web Conferences Steering Committee}, \bibinfo{address}{Republic and
  Canton of Geneva, CHE}, \bibinfo{pages}{531–541}.
\newblock
\showISBNx{9781450341431}
\urldef\tempurl%
\url{https://doi.org/10.1145/2872427.2883033}
\showDOI{\tempurl}


\bibitem[Borisov et~al\mbox{.}(2018)]%
        {csm}
\bibfield{author}{\bibinfo{person}{Alexey Borisov}, \bibinfo{person}{Martijn
  Wardenaar}, \bibinfo{person}{Ilya Markov}, {and} \bibinfo{person}{Maarten de
  Rijke}.} \bibinfo{year}{2018}\natexlab{}.
\newblock \showarticletitle{A Click Sequence Model for Web Search}. In
  \bibinfo{booktitle}{\emph{The 41st International ACM SIGIR Conference on
  Research and Development in Information Retrieval}} (Ann Arbor, MI, USA)
  \emph{(\bibinfo{series}{SIGIR '18})}. \bibinfo{publisher}{Association for
  Computing Machinery}, \bibinfo{address}{New York, NY, USA},
  \bibinfo{pages}{45–54}.
\newblock
\showISBNx{9781450356572}
\urldef\tempurl%
\url{https://doi.org/10.1145/3209978.3210004}
\showDOI{\tempurl}


\bibitem[Bruch et~al\mbox{.}(2020)]%
        {bruch-2020-pl}
\bibfield{author}{\bibinfo{person}{Sebastian Bruch}, \bibinfo{person}{Shuguang
  Han}, \bibinfo{person}{Michael Bendersky}, {and} \bibinfo{person}{Marc
  Najork}.} \bibinfo{year}{2020}\natexlab{}.
\newblock \showarticletitle{A Stochastic Treatment of Learning to Rank Scoring
  Functions}. In \bibinfo{booktitle}{\emph{Proceedings of the 13th
  International Conference on Web Search and Data Mining}} (Houston, TX, USA)
  \emph{(\bibinfo{series}{WSDM '20})}. \bibinfo{publisher}{Association for
  Computing Machinery}, \bibinfo{address}{New York, NY, USA},
  \bibinfo{pages}{61–69}.
\newblock
\showISBNx{9781450368223}
\urldef\tempurl%
\url{https://doi.org/10.1145/3336191.3371844}
\showDOI{\tempurl}


\bibitem[Burges(2010)]%
        {lambdamart}
\bibfield{author}{\bibinfo{person}{Christopher J.~C. Burges}.}
  \bibinfo{year}{2010}\natexlab{}.
\newblock \bibinfo{booktitle}{\emph{From {RankNet} to {LambdaRank} to
  {LambdaMART}: An Overview}}.
\newblock \bibinfo{type}{{T}echnical {R}eport}. \bibinfo{institution}{Microsoft
  Research}.
\newblock
\urldef\tempurl%
\url{http://research.microsoft.com/en-us/um/people/cburges/tech\_reports/MSR-TR-2010-82.pdf}
\showURL{%
\tempurl}


\bibitem[Chapelle and Chang(2011)]%
        {chapelle-2011-yahoo}
\bibfield{author}{\bibinfo{person}{Olivier Chapelle} {and} \bibinfo{person}{Yi
  Chang}.} \bibinfo{year}{2011}\natexlab{}.
\newblock \showarticletitle{Yahoo! Learning to Rank Challenge Overview}. In
  \bibinfo{booktitle}{\emph{Proceedings of the Learning to Rank Challenge}}
  \emph{(\bibinfo{series}{Proceedings of Machine Learning Research},
  Vol.~\bibinfo{volume}{14})}, \bibfield{editor}{\bibinfo{person}{Olivier
  Chapelle}, \bibinfo{person}{Yi~Chang}, {and} \bibinfo{person}{Tie-Yan Liu}}
  (Eds.). \bibinfo{publisher}{PMLR}, \bibinfo{address}{Haifa, Israel},
  \bibinfo{pages}{1--24}.
\newblock
\urldef\tempurl%
\url{https://proceedings.mlr.press/v14/chapelle11a.html}
\showURL{%
\tempurl}


\bibitem[Chapelle et~al\mbox{.}(2009)]%
        {err}
\bibfield{author}{\bibinfo{person}{Olivier Chapelle}, \bibinfo{person}{Donald
  Metlzer}, \bibinfo{person}{Ya Zhang}, {and} \bibinfo{person}{Pierre
  Grinspan}.} \bibinfo{year}{2009}\natexlab{}.
\newblock \showarticletitle{Expected Reciprocal Rank for Graded Relevance}. In
  \bibinfo{booktitle}{\emph{Proceedings of the 18th ACM Conference on
  Information and Knowledge Management}} (Hong Kong, China)
  \emph{(\bibinfo{series}{CIKM '09})}. \bibinfo{publisher}{Association for
  Computing Machinery}, \bibinfo{address}{New York, NY, USA},
  \bibinfo{pages}{621–630}.
\newblock
\showISBNx{9781605585123}
\urldef\tempurl%
\url{https://doi.org/10.1145/1645953.1646033}
\showDOI{\tempurl}


\bibitem[Chapelle and Zhang(2009)]%
        {dbn}
\bibfield{author}{\bibinfo{person}{Olivier Chapelle} {and} \bibinfo{person}{Ya
  Zhang}.} \bibinfo{year}{2009}\natexlab{}.
\newblock \showarticletitle{A Dynamic Bayesian Network Click Model for Web
  Search Ranking}. In \bibinfo{booktitle}{\emph{Proceedings of the 18th
  International Conference on World Wide Web}} (Madrid, Spain)
  \emph{(\bibinfo{series}{WWW '09})}. \bibinfo{publisher}{ACM},
  \bibinfo{address}{New York, NY, USA}, \bibinfo{pages}{1–10}.
\newblock
\showISBNx{9781605584874}
\urldef\tempurl%
\url{https://doi.org/10.1145/1526709.1526711}
\showDOI{\tempurl}


\bibitem[Chen et~al\mbox{.}(2020)]%
        {cacm}
\bibfield{author}{\bibinfo{person}{Jia Chen}, \bibinfo{person}{Jiaxin Mao},
  \bibinfo{person}{Yiqun Liu}, \bibinfo{person}{Min Zhang}, {and}
  \bibinfo{person}{Shaoping Ma}.} \bibinfo{year}{2020}\natexlab{}.
\newblock \showarticletitle{A Context-Aware Click Model for Web Search}. In
  \bibinfo{booktitle}{\emph{Proceedings of the 13th International Conference on
  Web Search and Data Mining}} (Houston, TX, USA) \emph{(\bibinfo{series}{WSDM
  '20})}. \bibinfo{publisher}{ACM}, \bibinfo{address}{New York, NY, USA},
  \bibinfo{pages}{88–96}.
\newblock
\showISBNx{9781450368223}
\urldef\tempurl%
\url{https://doi.org/10.1145/3336191.3371819}
\showDOI{\tempurl}


\bibitem[Chen and Yan(2012)]%
        {chen-2012-position}
\bibfield{author}{\bibinfo{person}{Ye Chen} {and} \bibinfo{person}{Tak~W.
  Yan}.} \bibinfo{year}{2012}\natexlab{}.
\newblock \showarticletitle{Position-Normalized Click Prediction in Search
  Advertising} \emph{(\bibinfo{series}{KDD '12})}.
  \bibinfo{publisher}{Association for Computing Machinery},
  \bibinfo{address}{New York, NY, USA}, \bibinfo{pages}{795–803}.
\newblock
\showISBNx{9781450314626}
\urldef\tempurl%
\url{https://doi.org/10.1145/2339530.2339654}
\showDOI{\tempurl}


\bibitem[Chuklin et~al\mbox{.}(2015)]%
        {cm_book}
\bibfield{author}{\bibinfo{person}{Aleksandr Chuklin}, \bibinfo{person}{Ilya
  Markov}, {and} \bibinfo{person}{Maarten de Rijke}.}
  \bibinfo{year}{2015}\natexlab{}.
\newblock \bibinfo{booktitle}{\emph{Click Models for Web Search}}.
\newblock \bibinfo{publisher}{Morgan \& Claypool}.
\newblock
\showISBNx{9781627056489}
\urldef\tempurl%
\url{https://doi.org/10.2200/S00654ED1V01Y201507ICR043}
\showDOI{\tempurl}


\bibitem[Chuklin et~al\mbox{.}(2013)]%
        {cm-offline-metrics}
\bibfield{author}{\bibinfo{person}{Aleksandr Chuklin}, \bibinfo{person}{Pavel
  Serdyukov}, {and} \bibinfo{person}{Maarten de Rijke}.}
  \bibinfo{year}{2013}\natexlab{}.
\newblock \showarticletitle{Click Model-Based Information Retrieval Metrics}.
  In \bibinfo{booktitle}{\emph{Proceedings of the 36th International ACM SIGIR
  Conference on Research and Development in Information Retrieval}} (Dublin,
  Ireland) \emph{(\bibinfo{series}{SIGIR '13})}.
  \bibinfo{publisher}{Association for Computing Machinery},
  \bibinfo{address}{New York, NY, USA}, \bibinfo{pages}{493–502}.
\newblock
\showISBNx{9781450320344}
\urldef\tempurl%
\url{https://doi.org/10.1145/2484028.2484071}
\showDOI{\tempurl}


\bibitem[Craswell et~al\mbox{.}(2008)]%
        {pbm}
\bibfield{author}{\bibinfo{person}{Nick Craswell}, \bibinfo{person}{Onno
  Zoeter}, \bibinfo{person}{Michael Taylor}, {and} \bibinfo{person}{Bill
  Ramsey}.} \bibinfo{year}{2008}\natexlab{}.
\newblock \showarticletitle{An Experimental Comparison of Click Position-Bias
  Models}. In \bibinfo{booktitle}{\emph{Proceedings of the 2008 International
  Conference on Web Search and Data Mining}} (Palo Alto, CA, USA)
  \emph{(\bibinfo{series}{WSDM '08})}. \bibinfo{publisher}{ACM},
  \bibinfo{address}{New York, NY, USA}, \bibinfo{pages}{87–94}.
\newblock
\showISBNx{9781595939272}
\urldef\tempurl%
\url{https://doi.org/10.1145/1341531.1341545}
\showDOI{\tempurl}


\bibitem[Dai et~al\mbox{.}(2021)]%
        {aicm}
\bibfield{author}{\bibinfo{person}{Xinyi Dai}, \bibinfo{person}{Jianghao Lin},
  \bibinfo{person}{Weinan Zhang}, \bibinfo{person}{Shuai Li},
  \bibinfo{person}{Weiwen Liu}, \bibinfo{person}{Ruiming Tang},
  \bibinfo{person}{Xiuqiang He}, \bibinfo{person}{Jianye Hao},
  \bibinfo{person}{Jun Wang}, {and} \bibinfo{person}{Yong Yu}.}
  \bibinfo{year}{2021}\natexlab{}.
\newblock \showarticletitle{An Adversarial Imitation Click Model for
  Information Retrieval}. In \bibinfo{booktitle}{\emph{Proceedings of the Web
  Conference 2021}} (Ljubljana, Slovenia) \emph{(\bibinfo{series}{WWW '21})}.
  \bibinfo{publisher}{ACM}, \bibinfo{address}{New York, NY, USA},
  \bibinfo{pages}{1809–1820}.
\newblock
\showISBNx{9781450383127}
\urldef\tempurl%
\url{https://doi.org/10.1145/3442381.3449913}
\showDOI{\tempurl}


\bibitem[Dawid(1979)]%
        {dawid-1979-ci}
\bibfield{author}{\bibinfo{person}{Alexander~Philip Dawid}.}
  \bibinfo{year}{1979}\natexlab{}.
\newblock \showarticletitle{Conditional Independence in Statistical Theory}.
\newblock \bibinfo{journal}{\emph{Journal of the Royal Statistical Society:
  Series B (Methodological)}} \bibinfo{volume}{41}, \bibinfo{number}{1}
  (\bibinfo{year}{1979}), \bibinfo{pages}{1--15}.
\newblock
\urldef\tempurl%
\url{https://doi.org/10.1111/j.2517-6161.1979.tb01052.x}
\showDOI{\tempurl}


\bibitem[Deffayet et~al\mbox{.}(2023)]%
        {deffayet-2022}
\bibfield{author}{\bibinfo{person}{Romain Deffayet},
  \bibinfo{person}{Jean-Michel Renders}, {and} \bibinfo{person}{Maarten de
  Rijke}.} \bibinfo{year}{2023}\natexlab{}.
\newblock \showarticletitle{Evaluating the Robustness of Click Models to Policy
  Distributional Shift}.
\newblock \bibinfo{journal}{\emph{ACM Trans. Inf. Syst.}} \bibinfo{volume}{41},
  \bibinfo{number}{4}, Article \bibinfo{articleno}{84} (\bibinfo{date}{mar}
  \bibinfo{year}{2023}), \bibinfo{numpages}{28}~pages.
\newblock
\showISSN{1046-8188}
\urldef\tempurl%
\url{https://doi.org/10.1145/3569086}
\showDOI{\tempurl}


\bibitem[Doran et~al\mbox{.}(2014)]%
        {doran-2014-permutation-kernel}
\bibfield{author}{\bibinfo{person}{Gary Doran}, \bibinfo{person}{Krikamol
  Muandet}, \bibinfo{person}{Kun Zhang}, {and} \bibinfo{person}{Bernhard
  Sch\"{o}lkopf}.} \bibinfo{year}{2014}\natexlab{}.
\newblock \showarticletitle{A Permutation-Based Kernel Conditional Independence
  Test}. In \bibinfo{booktitle}{\emph{Proceedings of the Thirtieth Conference
  on Uncertainty in Artificial Intelligence}} (Quebec City, Quebec, Canada)
  \emph{(\bibinfo{series}{UAI'14})}. \bibinfo{publisher}{AUAI Press},
  \bibinfo{address}{Arlington, Virginia, USA}, \bibinfo{pages}{132–141}.
\newblock
\showISBNx{9780974903910}


\bibitem[Dupret and Piwowarski(2008)]%
        {ubm}
\bibfield{author}{\bibinfo{person}{Georges~E. Dupret} {and}
  \bibinfo{person}{Benjamin Piwowarski}.} \bibinfo{year}{2008}\natexlab{}.
\newblock \showarticletitle{A User Browsing Model to Predict Search Engine
  Click Data from Past Observations.}. In \bibinfo{booktitle}{\emph{Proceedings
  of the 31st Annual International ACM SIGIR Conference on Research and
  Development in Information Retrieval}} (Singapore, Singapore)
  \emph{(\bibinfo{series}{SIGIR '08})}. \bibinfo{publisher}{ACM},
  \bibinfo{address}{New York, NY, USA}, \bibinfo{pages}{331–338}.
\newblock
\showISBNx{9781605581644}
\urldef\tempurl%
\url{https://doi.org/10.1145/1390334.1390392}
\showDOI{\tempurl}


\bibitem[Fukumizu et~al\mbox{.}(2004)]%
        {fukumizu-2004-kernel-hilbert}
\bibfield{author}{\bibinfo{person}{Kenji Fukumizu}, \bibinfo{person}{Francis~R.
  Bach}, {and} \bibinfo{person}{Michael~I. Jordan}.}
  \bibinfo{year}{2004}\natexlab{}.
\newblock \showarticletitle{Dimensionality Reduction for Supervised Learning
  with Reproducing Kernel Hilbert Spaces}.
\newblock \bibinfo{journal}{\emph{J. Mach. Learn. Res.}}  \bibinfo{volume}{5}
  (\bibinfo{date}{dec} \bibinfo{year}{2004}), \bibinfo{pages}{73–99}.
\newblock
\showISSN{1532-4435}
\urldef\tempurl%
\url{https://dl.acm.org/doi/10.5555/1005332.1005335}
\showURL{%
\tempurl}


\bibitem[Grotov et~al\mbox{.}(2015)]%
        {grotov-2015-comparative}
\bibfield{author}{\bibinfo{person}{Artem Grotov}, \bibinfo{person}{Aleksandr
  Chuklin}, \bibinfo{person}{Ilya Markov}, \bibinfo{person}{Luka Stout},
  \bibinfo{person}{Finde Xumara}, {and} \bibinfo{person}{Maarten de Rijke}.}
  \bibinfo{year}{2015}\natexlab{}.
\newblock \showarticletitle{A Comparative Study of Click Models for Web
  Search}. In \bibinfo{booktitle}{\emph{Proceedings of the 6th International
  Conference on Experimental IR Meets Multilinguality, Multimodality, and
  Interaction - Volume 9283}} (Toulouse, France)
  \emph{(\bibinfo{series}{CLEF'15})}. \bibinfo{publisher}{Springer-Verlag},
  \bibinfo{address}{Berlin, Heidelberg}, \bibinfo{pages}{78–90}.
\newblock
\showISBNx{9783319240268}
\urldef\tempurl%
\url{https://doi.org/10.1007/978-3-319-24027-5_7}
\showDOI{\tempurl}


\bibitem[Guo et~al\mbox{.}(2021)]%
        {debiasing-conversion}
\bibfield{author}{\bibinfo{person}{Siyuan Guo}, \bibinfo{person}{Lixin Zou},
  \bibinfo{person}{Yiding Liu}, \bibinfo{person}{Wenwen Ye},
  \bibinfo{person}{Suqi Cheng}, \bibinfo{person}{Shuaiqiang Wang},
  \bibinfo{person}{Hechang Chen}, \bibinfo{person}{Dawei Yin}, {and}
  \bibinfo{person}{Yi Chang}.} \bibinfo{year}{2021}\natexlab{}.
\newblock \showarticletitle{Enhanced Doubly Robust Learning for Debiasing
  Post-Click Conversion Rate Estimation}. In
  \bibinfo{booktitle}{\emph{Proceedings of the 44th International ACM SIGIR
  Conference on Research and Development in Information Retrieval}} (Virtual
  Event, Canada) \emph{(\bibinfo{series}{SIGIR '21})}.
  \bibinfo{publisher}{Association for Computing Machinery},
  \bibinfo{address}{New York, NY, USA}, \bibinfo{pages}{275–284}.
\newblock
\showISBNx{9781450380379}
\urldef\tempurl%
\url{https://doi.org/10.1145/3404835.3462917}
\showDOI{\tempurl}


\bibitem[Huang et~al\mbox{.}(2020)]%
        {sofa}
\bibfield{author}{\bibinfo{person}{Jin Huang}, \bibinfo{person}{Harrie
  Oosterhuis}, \bibinfo{person}{Maarten de Rijke}, {and} \bibinfo{person}{Herke
  van Hoof}.} \bibinfo{year}{2020}\natexlab{}.
\newblock \showarticletitle{Keeping Dataset Biases out of the Simulation: A
  Debiased Simulator for Reinforcement Learning based Recommender Systems}. In
  \bibinfo{booktitle}{\emph{Proceedings of the 14th ACM Conference on
  Recommender Systems}} (Virtual Event, Brazil) \emph{(\bibinfo{series}{RecSys
  '20})}. \bibinfo{publisher}{Association for Computing Machinery},
  \bibinfo{address}{New York, NY, USA}, \bibinfo{pages}{190–199}.
\newblock
\showISBNx{9781450375832}
\urldef\tempurl%
\url{https://doi.org/10.1145/3383313.3412252}
\showDOI{\tempurl}


\bibitem[Joachims et~al\mbox{.}(2005)]%
        {joachims-2005-position-bias}
\bibfield{author}{\bibinfo{person}{Thorsten Joachims}, \bibinfo{person}{Laura
  Granka}, \bibinfo{person}{Bing Pan}, \bibinfo{person}{Helene Hembrooke},
  {and} \bibinfo{person}{Geri Gay}.} \bibinfo{year}{2005}\natexlab{}.
\newblock \showarticletitle{Accurately Interpreting Clickthrough Data as
  Implicit Feedback}. In \bibinfo{booktitle}{\emph{Proceedings of the 28th
  Annual International ACM SIGIR Conference on Research and Development in
  Information Retrieval}} (Salvador, Brazil) \emph{(\bibinfo{series}{SIGIR
  '05})}. \bibinfo{publisher}{Association for Computing Machinery},
  \bibinfo{address}{New York, NY, USA}, \bibinfo{pages}{154–161}.
\newblock
\showISBNx{1595930345}
\urldef\tempurl%
\url{https://doi.org/10.1145/1076034.1076063}
\showDOI{\tempurl}


\bibitem[Joachims et~al\mbox{.}(2017)]%
        {unbiased_joachims}
\bibfield{author}{\bibinfo{person}{Thorsten Joachims}, \bibinfo{person}{Adith
  Swaminathan}, {and} \bibinfo{person}{Tobias Schnabel}.}
  \bibinfo{year}{2017}\natexlab{}.
\newblock \showarticletitle{Unbiased Learning-to-Rank with Biased Feedback}. In
  \bibinfo{booktitle}{\emph{Proceedings of the Tenth ACM International
  Conference on Web Search and Data Mining}} (Cambridge, United Kingdom)
  \emph{(\bibinfo{series}{WSDM '17})}. \bibinfo{publisher}{Association for
  Computing Machinery}, \bibinfo{address}{New York, NY, USA},
  \bibinfo{pages}{781–789}.
\newblock
\showISBNx{9781450346757}
\urldef\tempurl%
\url{https://doi.org/10.1145/3018661.3018699}
\showDOI{\tempurl}


\bibitem[Koller and Friedman(2009)]%
        {koller-2009-pgm}
\bibfield{author}{\bibinfo{person}{Daphne Koller} {and} \bibinfo{person}{Nir
  Friedman}.} \bibinfo{year}{2009}\natexlab{}.
\newblock \bibinfo{booktitle}{\emph{Probabilistic graphical models: principles
  and techniques}}.
\newblock \bibinfo{publisher}{MIT press}.
\newblock
\urldef\tempurl%
\url{https://mitpress.mit.edu/9780262013192/probabilistic-graphical-models/}
\showURL{%
\tempurl}


\bibitem[Koller and Sahami(1996)]%
        {koller-1996-toward}
\bibfield{author}{\bibinfo{person}{Daphne Koller} {and} \bibinfo{person}{Mehran
  Sahami}.} \bibinfo{year}{1996}\natexlab{}.
\newblock \showarticletitle{Toward Optimal Feature Selection}. In
  \bibinfo{booktitle}{\emph{Proceedings of the Thirteenth International
  Conference on International Conference on Machine Learning}} (Bari, Italy)
  \emph{(\bibinfo{series}{ICML'96})}. \bibinfo{publisher}{Morgan Kaufmann
  Publishers Inc.}, \bibinfo{address}{San Francisco, CA, USA},
  \bibinfo{pages}{284–292}.
\newblock
\showISBNx{1558604197}
\urldef\tempurl%
\url{https://dl.acm.org/doi/10.5555/3091696.3091731}
\showURL{%
\tempurl}


\bibitem[Lin et~al\mbox{.}(2021)]%
        {graphcm}
\bibfield{author}{\bibinfo{person}{Jianghao Lin}, \bibinfo{person}{Weiwen Liu},
  \bibinfo{person}{Xinyi Dai}, \bibinfo{person}{Weinan Zhang},
  \bibinfo{person}{Shuai Li}, \bibinfo{person}{Ruiming Tang},
  \bibinfo{person}{Xiuqiang He}, \bibinfo{person}{Jianye Hao}, {and}
  \bibinfo{person}{Yong Yu}.} \bibinfo{year}{2021}\natexlab{}.
\newblock \showarticletitle{A Graph-Enhanced Click Model for Web Search}. In
  \bibinfo{booktitle}{\emph{Proceedings of the 44th International ACM SIGIR
  Conference on Research and Development in Information Retrieval}} (Virtual
  Event, Canada) \emph{(\bibinfo{series}{SIGIR '21})}.
  \bibinfo{publisher}{Association for Computing Machinery},
  \bibinfo{address}{New York, NY, USA}, \bibinfo{pages}{1259–1268}.
\newblock
\showISBNx{9781450380379}
\urldef\tempurl%
\url{https://doi.org/10.1145/3404835.3462895}
\showDOI{\tempurl}


\bibitem[Liu(2009)]%
        {liu-2009-ltr}
\bibfield{author}{\bibinfo{person}{Tie-Yan Liu}.}
  \bibinfo{year}{2009}\natexlab{}.
\newblock \showarticletitle{Learning to Rank for Information Retrieval}.
\newblock \bibinfo{journal}{\emph{Found. Trends Inf. Retr.}}
  \bibinfo{volume}{3}, \bibinfo{number}{3} (\bibinfo{date}{mar}
  \bibinfo{year}{2009}), \bibinfo{pages}{225–331}.
\newblock
\showISSN{1554-0669}
\urldef\tempurl%
\url{https://doi.org/10.1561/1500000016}
\showDOI{\tempurl}


\bibitem[Liu et~al\mbox{.}(2016)]%
        {TACM}
\bibfield{author}{\bibinfo{person}{Yiqun Liu}, \bibinfo{person}{Xiaohui Xie},
  \bibinfo{person}{Chao Wang}, \bibinfo{person}{Jian-Yun Nie},
  \bibinfo{person}{Min Zhang}, {and} \bibinfo{person}{Shaoping Ma}.}
  \bibinfo{year}{2016}\natexlab{}.
\newblock \showarticletitle{Time-Aware Click Model}.
\newblock \bibinfo{journal}{\emph{ACM Trans. Inf. Syst.}} \bibinfo{volume}{35},
  \bibinfo{number}{3}, Article \bibinfo{articleno}{16} (\bibinfo{date}{dec}
  \bibinfo{year}{2016}), \bibinfo{numpages}{24}~pages.
\newblock
\showISSN{1046-8188}
\urldef\tempurl%
\url{https://doi.org/10.1145/2988230}
\showDOI{\tempurl}


\bibitem[Lopez-Paz and Oquab(2017)]%
        {lopez-2016-classifier-two-sample}
\bibfield{author}{\bibinfo{person}{David Lopez-Paz} {and}
  \bibinfo{person}{Maxime Oquab}.} \bibinfo{year}{2017}\natexlab{}.
\newblock \showarticletitle{Revisiting Classifier Two-Sample Tests}. In
  \bibinfo{booktitle}{\emph{International Conference on Learning
  Representations}}.
\newblock
\urldef\tempurl%
\url{https://openreview.net/forum?id=SJkXfE5xx}
\showURL{%
\tempurl}


\bibitem[Luce(1959)]%
        {luce}
\bibfield{author}{\bibinfo{person}{Duncan Luce}.}
  \bibinfo{year}{1959}\natexlab{}.
\newblock \bibinfo{booktitle}{\emph{Individual Choice Behavior: A Theoretical
  Analysis}}.
\newblock \bibinfo{publisher}{Courier Corporation}.
\newblock
\urldef\tempurl%
\url{https://psycnet.apa.org/record/1960-03588-000}
\showURL{%
\tempurl}


\bibitem[Margaritis(2005)]%
        {margaritis-2005-discrete}
\bibfield{author}{\bibinfo{person}{Dimitris Margaritis}.}
  \bibinfo{year}{2005}\natexlab{}.
\newblock \showarticletitle{Distribution-Free Learning of Bayesian Network
  Structure in Continuous Domains}. In \bibinfo{booktitle}{\emph{Proceedings of
  the 20th National Conference on Artificial Intelligence - Volume 2}}
  (Pittsburgh, Pennsylvania) \emph{(\bibinfo{series}{AAAI'05})}.
  \bibinfo{publisher}{AAAI Press}, \bibinfo{pages}{825–830}.
\newblock
\showISBNx{157735236x}
\urldef\tempurl%
\url{https://dl.acm.org/doi/10.5555/1619410.1619465}
\showURL{%
\tempurl}


\bibitem[McMahan et~al\mbox{.}(2013)]%
        {mcmahan-2013-ad}
\bibfield{author}{\bibinfo{person}{H.~Brendan McMahan}, \bibinfo{person}{Gary
  Holt}, \bibinfo{person}{D. Sculley}, \bibinfo{person}{Michael Young},
  \bibinfo{person}{Dietmar Ebner}, \bibinfo{person}{Julian Grady},
  \bibinfo{person}{Lan Nie}, \bibinfo{person}{Todd Phillips},
  \bibinfo{person}{Eugene Davydov}, \bibinfo{person}{Daniel Golovin},
  \bibinfo{person}{Sharat Chikkerur}, \bibinfo{person}{Dan Liu},
  \bibinfo{person}{Martin Wattenberg}, \bibinfo{person}{Arnar~Mar
  Hrafnkelsson}, \bibinfo{person}{Tom Boulos}, {and} \bibinfo{person}{Jeremy
  Kubica}.} \bibinfo{year}{2013}\natexlab{}.
\newblock \showarticletitle{Ad Click Prediction: A View from the Trenches}. In
  \bibinfo{booktitle}{\emph{Proceedings of the 19th ACM SIGKDD International
  Conference on Knowledge Discovery and Data Mining}} (Chicago, Illinois, USA)
  \emph{(\bibinfo{series}{KDD '13})}. \bibinfo{publisher}{Association for
  Computing Machinery}, \bibinfo{address}{New York, NY, USA},
  \bibinfo{pages}{1222–1230}.
\newblock
\showISBNx{9781450321747}
\urldef\tempurl%
\url{https://doi.org/10.1145/2487575.2488200}
\showDOI{\tempurl}


\bibitem[Mukherjee et~al\mbox{.}(2019)]%
        {mukherjee-2020-ccmi}
\bibfield{author}{\bibinfo{person}{Sudipto Mukherjee},
  \bibinfo{person}{Himanshu Asnani}, {and} \bibinfo{person}{Sreeram Kannan}.}
  \bibinfo{year}{2019}\natexlab{}.
\newblock \showarticletitle{{CCMI} : Classifier based Conditional Mutual
  Information Estimation}. In \bibinfo{booktitle}{\emph{Proceedings of the
  Thirty-Fifth Conference on Uncertainty in Artificial Intelligence, (UAI)}}
  \emph{(\bibinfo{series}{Proceedings of Machine Learning Research},
  Vol.~\bibinfo{volume}{115})}. \bibinfo{publisher}{{AUAI} Press},
  \bibinfo{pages}{1083--1093}.
\newblock
\urldef\tempurl%
\url{http://proceedings.mlr.press/v115/mukherjee20a.html}
\showURL{%
\tempurl}


\bibitem[Oosterhuis(2021)]%
        {harrie_plopt}
\bibfield{author}{\bibinfo{person}{Harrie Oosterhuis}.}
  \bibinfo{year}{2021}\natexlab{}.
\newblock \showarticletitle{Computationally Efficient Optimization of
  Plackett-Luce Ranking Models for Relevance and Fairness}. In
  \bibinfo{booktitle}{\emph{Proceedings of the 44th International ACM SIGIR
  Conference on Research and Development in Information Retrieval}}.
  \bibinfo{publisher}{Association for Computing Machinery},
  \bibinfo{address}{New York, NY, USA}, \bibinfo{pages}{1023–1032}.
\newblock
\showISBNx{9781450380379}
\urldef\tempurl%
\url{https://doi.org/10.1145/3404835.3462830}
\showURL{%
\tempurl}


\bibitem[Oosterhuis(2022)]%
        {harrie-unbiasedness}
\bibfield{author}{\bibinfo{person}{Harrie Oosterhuis}.}
  \bibinfo{year}{2022}\natexlab{}.
\newblock \showarticletitle{Reaching the End of Unbiasedness: Uncovering
  Implicit Limitations of Click-Based Learning to Rank}. In
  \bibinfo{booktitle}{\emph{Proceedings of the 2022 ACM SIGIR International
  Conference on Theory of Information Retrieval}} (Madrid, Spain)
  \emph{(\bibinfo{series}{ICTIR '22})}. \bibinfo{publisher}{Association for
  Computing Machinery}, \bibinfo{address}{New York, NY, USA},
  \bibinfo{pages}{264–274}.
\newblock
\showISBNx{9781450394123}
\urldef\tempurl%
\url{https://doi.org/10.1145/3539813.3545137}
\showDOI{\tempurl}


\bibitem[Oosterhuis and de~Rijke(2020a)]%
        {policy-aware}
\bibfield{author}{\bibinfo{person}{Harrie Oosterhuis} {and}
  \bibinfo{person}{Maarten de Rijke}.} \bibinfo{year}{2020}\natexlab{a}.
\newblock \bibinfo{booktitle}{\emph{Policy-Aware Unbiased Learning to Rank for
  Top-k Rankings}}.
\newblock \bibinfo{publisher}{Association for Computing Machinery},
  \bibinfo{address}{New York, NY, USA}, \bibinfo{pages}{489–498}.
\newblock
\showISBNx{9781450380164}
\urldef\tempurl%
\url{https://doi.org/10.1145/3397271.3401102}
\showURL{%
\tempurl}


\bibitem[Oosterhuis and de~Rijke(2020b)]%
        {oosterhuis-2020-taking}
\bibfield{author}{\bibinfo{person}{Harrie Oosterhuis} {and}
  \bibinfo{person}{Maarten de Rijke}.} \bibinfo{year}{2020}\natexlab{b}.
\newblock \showarticletitle{Taking the Counterfactual Online: Efficient and
  Unbiased Online Evaluation for Ranking}. In
  \bibinfo{booktitle}{\emph{Proceedings of the 2020 ACM SIGIR on International
  Conference on Theory of Information Retrieval}} (Virtual Event, Norway)
  \emph{(\bibinfo{series}{ICTIR '20})}. \bibinfo{publisher}{Association for
  Computing Machinery}, \bibinfo{address}{New York, NY, USA},
  \bibinfo{pages}{137–144}.
\newblock
\showISBNx{9781450380676}
\urldef\tempurl%
\url{https://doi.org/10.1145/3409256.3409820}
\showDOI{\tempurl}


\bibitem[Ovaisi et~al\mbox{.}(2020)]%
        {ovaisi-2020-selection-bias}
\bibfield{author}{\bibinfo{person}{Zohreh Ovaisi}, \bibinfo{person}{Ragib
  Ahsan}, \bibinfo{person}{Yifan Zhang}, \bibinfo{person}{Kathryn Vasilaky},
  {and} \bibinfo{person}{Elena Zheleva}.} \bibinfo{year}{2020}\natexlab{}.
\newblock \showarticletitle{Correcting for Selection Bias in Learning-to-Rank
  Systems}. In \bibinfo{booktitle}{\emph{Proceedings of The Web Conference
  2020}} (Taipei, Taiwan) \emph{(\bibinfo{series}{WWW '20})}.
  \bibinfo{publisher}{Association for Computing Machinery},
  \bibinfo{address}{New York, NY, USA}, \bibinfo{pages}{1863–1873}.
\newblock
\showISBNx{9781450370233}
\urldef\tempurl%
\url{https://doi.org/10.1145/3366423.3380255}
\showDOI{\tempurl}


\bibitem[Pearl(2009)]%
        {pearl-2009-causality}
\bibfield{author}{\bibinfo{person}{Judea Pearl}.}
  \bibinfo{year}{2009}\natexlab{}.
\newblock \bibinfo{booktitle}{\emph{Causality}}.
\newblock \bibinfo{publisher}{Cambridge university press}.
\newblock
\urldef\tempurl%
\url{https://doi.org/10.1017/CBO9780511803161}
\showDOI{\tempurl}


\bibitem[Plackett(1975)]%
        {plackett}
\bibfield{author}{\bibinfo{person}{Robin Plackett}.}
  \bibinfo{year}{1975}\natexlab{}.
\newblock \showarticletitle{The Analysis of Permutations}.
\newblock \bibinfo{journal}{\emph{Journal of the Royal Statistical Society}}
  \bibinfo{volume}{24}, \bibinfo{number}{2} (\bibinfo{year}{1975}).
\newblock
\urldef\tempurl%
\url{https://doi.org/10.2307/2346567}
\showDOI{\tempurl}


\bibitem[Qin and Liu(2013)]%
        {mslr}
\bibfield{author}{\bibinfo{person}{Tao Qin} {and} \bibinfo{person}{Tie{-}Yan
  Liu}.} \bibinfo{year}{2013}\natexlab{}.
\newblock \bibinfo{title}{Introducing {LETOR} 4.0 Datasets}.
\newblock
\newblock
\urldef\tempurl%
\url{https://arxiv.org/abs/1306.2597}
\showURL{%
\tempurl}


\bibitem[Rahdari et~al\mbox{.}(2022)]%
        {rahdari2022carousel}
\bibfield{author}{\bibinfo{person}{Behnam Rahdari}, \bibinfo{person}{Branislav
  Kveton}, {and} \bibinfo{person}{Peter Brusilovsky}.}
  \bibinfo{year}{2022}\natexlab{}.
\newblock \bibinfo{title}{From Ranked Lists to Carousels: A Carousel Click
  Model}.
\newblock
\newblock
\urldef\tempurl%
\url{https://arxiv.org/abs/2209.13426}
\showURL{%
\tempurl}


\bibitem[Sen et~al\mbox{.}(2017)]%
        {sen-2017-ccit}
\bibfield{author}{\bibinfo{person}{Rajat Sen}, \bibinfo{person}{Ananda~Theertha
  Suresh}, \bibinfo{person}{Karthikeyan Shanmugam},
  \bibinfo{person}{Alexandres~G. Dimakis}, {and} \bibinfo{person}{Sanjay
  Shakkettai}.} \bibinfo{year}{2017}\natexlab{}.
\newblock \showarticletitle{Model-Powered Conditional Independence Test}. In
  \bibinfo{booktitle}{\emph{Proceedings of the 31st International Conference on
  Neural Information Processing Systems}} (Long Beach, California, USA)
  \emph{(\bibinfo{series}{NIPS'17})}. \bibinfo{publisher}{Curran Associates
  Inc.}, \bibinfo{address}{Red Hook, NY, USA}, \bibinfo{pages}{2955–2965}.
\newblock
\showISBNx{9781510860964}
\urldef\tempurl%
\url{https://dl.acm.org/doi/10.5555/3294996.3295055}
\showURL{%
\tempurl}


\bibitem[Serdyukov et~al\mbox{.}(2012)]%
        {yandex}
\bibfield{author}{\bibinfo{person}{Pavel Serdyukov}, \bibinfo{person}{Nick
  Craswell}, {and} \bibinfo{person}{Georges Dupret}.}
  \bibinfo{year}{2012}\natexlab{}.
\newblock \showarticletitle{WSCD 2012: Workshop on Web Search Click Data 2012}.
  In \bibinfo{booktitle}{\emph{Proceedings of the Fifth ACM International
  Conference on Web Search and Data Mining}} (Seattle, Washington, USA)
  \emph{(\bibinfo{series}{WSDM '12})}. \bibinfo{publisher}{ACM},
  \bibinfo{address}{New York, NY, USA}, \bibinfo{pages}{771–772}.
\newblock
\showISBNx{9781450307475}
\urldef\tempurl%
\url{https://doi.org/10.1145/2124295.2124396}
\showDOI{\tempurl}


\bibitem[Shen et~al\mbox{.}(2021)]%
        {ood-generalization}
\bibfield{author}{\bibinfo{person}{Zheyan Shen}, \bibinfo{person}{Jiashuo Liu},
  \bibinfo{person}{Yue He}, \bibinfo{person}{Xingxuan Zhang},
  \bibinfo{person}{Renzhe Xu}, \bibinfo{person}{Han Yu}, {and}
  \bibinfo{person}{Peng Cui}.} \bibinfo{year}{2021}\natexlab{}.
\newblock \bibinfo{title}{Towards Out-Of-Distribution Generalization: A
  Survey}.
\newblock
\newblock
\urldef\tempurl%
\url{https://doi.org/10.48550/ARXIV.2108.13624}
\showDOI{\tempurl}


\bibitem[Slivkins(2019)]%
        {slivkins-2019-introduction}
\bibfield{author}{\bibinfo{person}{Aleksandrs Slivkins}.}
  \bibinfo{year}{2019}\natexlab{}.
\newblock \showarticletitle{Introduction to multi-armed bandits}.
\newblock \bibinfo{journal}{\emph{Foundations and Trends in Machine Learning}}
  \bibinfo{volume}{12}, \bibinfo{number}{1-2} (\bibinfo{year}{2019}),
  \bibinfo{pages}{1--286}.
\newblock
\urldef\tempurl%
\url{https://doi.org/10.1561/2200000068}
\showDOI{\tempurl}


\bibitem[Su and White(2007)]%
        {su-2007-ci}
\bibfield{author}{\bibinfo{person}{Liangjun Su} {and} \bibinfo{person}{Halbert
  White}.} \bibinfo{year}{2007}\natexlab{}.
\newblock \showarticletitle{A consistent characteristic function-based test for
  conditional independence}.
\newblock \bibinfo{journal}{\emph{Journal of Econometrics}}
  \bibinfo{volume}{141}, \bibinfo{number}{2} (\bibinfo{year}{2007}),
  \bibinfo{pages}{807--834}.
\newblock
\urldef\tempurl%
\url{https://doi.org/10.1016/j.jeconom.2006.11.006}
\showDOI{\tempurl}


\bibitem[Vardasbi et~al\mbox{.}(2020a)]%
        {cm-ips}
\bibfield{author}{\bibinfo{person}{Ali Vardasbi}, \bibinfo{person}{Maarten de
  Rijke}, {and} \bibinfo{person}{Ilya Markov}.}
  \bibinfo{year}{2020}\natexlab{a}.
\newblock \showarticletitle{Cascade Model-Based Propensity Estimation for
  Counterfactual Learning to Rank}. In \bibinfo{booktitle}{\emph{Proceedings of
  the 43rd International ACM SIGIR Conference on Research and Development in
  Information Retrieval}} (Virtual Event, China) \emph{(\bibinfo{series}{SIGIR
  '20})}. \bibinfo{publisher}{Association for Computing Machinery},
  \bibinfo{address}{New York, NY, USA}, \bibinfo{pages}{2089–2092}.
\newblock
\showISBNx{9781450380164}
\urldef\tempurl%
\url{https://doi.org/10.1145/3397271.3401299}
\showDOI{\tempurl}


\bibitem[Vardasbi et~al\mbox{.}(2020b)]%
        {affine-estimator}
\bibfield{author}{\bibinfo{person}{Ali Vardasbi}, \bibinfo{person}{Harrie
  Oosterhuis}, {and} \bibinfo{person}{Maarten de Rijke}.}
  \bibinfo{year}{2020}\natexlab{b}.
\newblock \showarticletitle{When Inverse Propensity Scoring Does Not Work:
  Affine Corrections for Unbiased Learning to Rank}. In
  \bibinfo{booktitle}{\emph{Proceedings of the 29th ACM International
  Conference on Information and Knowledge Management}} (Virtual Event, Ireland)
  \emph{(\bibinfo{series}{CIKM '20})}. \bibinfo{publisher}{Association for
  Computing Machinery}, \bibinfo{address}{New York, NY, USA},
  \bibinfo{pages}{1475–1484}.
\newblock
\showISBNx{9781450368599}
\urldef\tempurl%
\url{https://doi.org/10.1145/3340531.3412031}
\showDOI{\tempurl}


\bibitem[Zheng et~al\mbox{.}(2019)]%
        {mcm}
\bibfield{author}{\bibinfo{person}{Yukun Zheng}, \bibinfo{person}{Jiaxin Mao},
  \bibinfo{person}{Yiqun Liu}, \bibinfo{person}{Cheng Luo},
  \bibinfo{person}{Min Zhang}, {and} \bibinfo{person}{Shaoping Ma}.}
  \bibinfo{year}{2019}\natexlab{}.
\newblock \showarticletitle{Constructing Click Model for Mobile Search with
  Viewport Time}.
\newblock \bibinfo{journal}{\emph{ACM Trans. Inf. Syst.}} \bibinfo{volume}{37},
  \bibinfo{number}{4} (\bibinfo{date}{sep} \bibinfo{year}{2019}),
  \bibinfo{numpages}{34}~pages.
\newblock
\showISSN{1046-8188}
\urldef\tempurl%
\url{https://doi.org/10.1145/3360486}
\showDOI{\tempurl}


\bibitem[Zhu et~al\mbox{.}(2010)]%
        {zhu-2010-novel-click-model}
\bibfield{author}{\bibinfo{person}{Zeyuan~Allen Zhu}, \bibinfo{person}{Weizhu
  Chen}, \bibinfo{person}{Tom Minka}, \bibinfo{person}{Chenguang Zhu}, {and}
  \bibinfo{person}{Zheng Chen}.} \bibinfo{year}{2010}\natexlab{}.
\newblock \showarticletitle{A Novel Click Model and Its Applications to Online
  Advertising}. In \bibinfo{booktitle}{\emph{Proceedings of the Third ACM
  International Conference on Web Search and Data Mining}} (New York, New York,
  USA) \emph{(\bibinfo{series}{WSDM '10})}. \bibinfo{publisher}{Association for
  Computing Machinery}, \bibinfo{address}{New York, NY, USA},
  \bibinfo{pages}{321–330}.
\newblock
\showISBNx{9781605588896}
\urldef\tempurl%
\url{https://doi.org/10.1145/1718487.1718528}
\showDOI{\tempurl}


\bibitem[Zhuang et~al\mbox{.}(2021)]%
        {xpa}
\bibfield{author}{\bibinfo{person}{Honglei Zhuang}, \bibinfo{person}{Zhen Qin},
  \bibinfo{person}{Xuanhui Wang}, \bibinfo{person}{Michael Bendersky},
  \bibinfo{person}{Xinyu Qian}, \bibinfo{person}{Po Hu}, {and}
  \bibinfo{person}{Dan~Chary Chen}.} \bibinfo{year}{2021}\natexlab{}.
\newblock \showarticletitle{Cross-Positional Attention for Debiasing Clicks}.
  In \bibinfo{booktitle}{\emph{Proceedings of the Web Conference 2021}}
  (Ljubljana, Slovenia) \emph{(\bibinfo{series}{WWW '21})}.
  \bibinfo{publisher}{Association for Computing Machinery},
  \bibinfo{address}{New York, NY, USA}, \bibinfo{pages}{788–797}.
\newblock
\showISBNx{9781450383127}
\urldef\tempurl%
\url{https://doi.org/10.1145/3442381.3450098}
\showDOI{\tempurl}


\end{thebibliography}

\section{Appendix}
\appendix
\label{sec:appendix}

\section{Proof of Theorem~\ref{theorem-invariant-debiased}}
\label{app:proof}

\begin{proof}
First, a model such that $\tilde{r}^\mathcal{D}=0$ is debiased since $0 \Perp R_l \mid  (R, \mathcal{D})$, and therefore satisfies Eq.~\ref{invariant-implies-debiased}. Next, assume $\tilde{r}^\mathcal{D} \neq 0$ in the remainder. For a model following the examination-hypothesis, we can write, for any rank $k$, ranking $y$ and training dataset $\mathcal{D}$:
\begin{equation*}
c^\mathcal{D}(y)[k] = r^\mathcal{D}(y[k]) \times o^\mathcal{D}(y)[k].
\end{equation*}
Consider two documents $d_1$ and $d_2$ with $\tilde{r}^\mathcal{D}(d_2) \neq 0$. Consider also two rankings $y_1, y_2$ that differ only by their first document: $y_1[1] = d_1$, $y_2[1] = d_2$, and $y_1[k] = y_2[k]$ for $k > 1$. Let these rankings share the same examination probability on their first position: $\tilde{o}^\mathcal{D}(y_1)[1]= \tilde{o}^\mathcal{D}(y_2)[1]$, then we can write down the two click probabilities:
$$
\left\{\begin{matrix}
\tilde{c}^\mathcal{D}(y_1)[1] = \tilde{r}^\mathcal{D}(d_1) \times \tilde{o}^\mathcal{D}(y_1)[1]
\\ 
\tilde{c}^\mathcal{D}(y_2)[1] = \tilde{r}^\mathcal{D}(d_2) \times \tilde{o}^\mathcal{D}(y_2)[1]
\end{matrix}\right. 
$$
Because of the equality of examination probabilities, we have:
$$
\frac{\tilde{r}^\mathcal{D}(d_1)}{\tilde{r}^\mathcal{D}(d_2)} = \frac{\tilde{c}^\mathcal{D}(y_1)[1]}{\tilde{c}^\mathcal{D}(y_2)[1]} \overset{\textup{(LHS)}}{=} \frac{\tilde{c}(y_1)[1]}{\tilde{c}(y_2)[1]}
$$
When the left-hand side of Eq.~\ref{invariant-implies-debiased} is true, this ratio does not depend on $\pi_l$ and relevance scores are determined up to a document-independent constant, so knowing $R_l$ does not help predict the relevance of a newly picked document: $P(\tilde{R}^\mathcal{D} \mid \mathcal{D}, R) = P(\tilde{R}^\mathcal{D} \mid \mathcal{D}, R, R_l)$. Thus, the model is strongly debiasing: $\tilde{R}^\mathcal{D} \Perp R_l \mid (R, \mathcal{D})$.
\end{proof}

\noindent%
A key observation is that the estimated relevance of a document may depend on the training dataset and ultimately on the logging policy,  yet revealing $R_l \mid (R, \mathcal{D})$ might not help to predict $\tilde{R}^\mathcal{D} \mid (R, \mathcal{D})$. 
Take, for example, the \acs{PBM} model. Without explicit constraints, we can scale its inferred relevance scores up or down, and by adjusting the examination scores accordingly, the click probabilities stay constant. Then a perfectly fitted \acs{PBM} may be invariant under policy shift, while the exact set of parameters it recovers depends on the training dataset. However, for a fixed dataset, knowing $R_l$ does not help to predict $\tilde{R}^\mathcal{D}$ in this setting.

\end{document}